\documentclass[11pt,a4paper]{article} 
\pdfoutput=1
\usepackage{jcappub} 

\usepackage{graphicx}
\usepackage{amssymb}

\newcommand{\bn}{{\mathbf n}}

\newcommand{\al}{\alpha}

\newcommand{\de}{\delta}
\newcommand{\De}{\Delta}

\newcommand{\La}{\Lambda}
\newcommand{\la}{\lambda}
\newcommand{\Om}{\Omega}

\newcommand{\si}{\sigma}

\newcommand{\ra}{\rightarrow}

\newcommand{\be}{\begin{equation}}
\newcommand{\ee}{\end{equation}}

\newcommand{\bea}{\begin{eqnarray}}
\newcommand{\eea}{\end{eqnarray}}
\newcommand{\bean}{\begin{eqnarray*}}
\newcommand{\eean}{\end{eqnarray*}}

\newcommand{\mr}{\mathrm}

\newcommand{\lan}{\langle}
\newcommand{\ran}{\rangle}

\newcommand{\ring}{{\mathrm{ring}}}
\newcommand{\cmb}{{\mathrm{cmb}}}
\newcommand{\Max}{{\mathrm{max}}}
\newcommand{\loc}{{\mathrm{loc}}}
\newcommand{\ps}{{\mathrm{ps}}}
\newcommand{\sr}{{\mathrm{sr}}}
\newcommand{\nside}{{\mathrm{nside}}}
\newcommand{\pix}{{\mathrm{pix}}}

\notoc 

\begin{document}

\title{The Kolmogorov-Smirnov test for the CMB}

\author[1,2]{Mona Frommert,}
\author[1]{Ruth Durrer}
\author[3]{and J\'er\^ome Michaud}

\affiliation[1]{D\'epartement de Physique Th\'eorique and Center for Astroparticle Physics, \\
Universit\'e de Gen\`eve, 24 quai Ernest Ansermet, CH--1211 Gen\`eve 4, Switzerland}
\affiliation[2]{Max-Planck-Institut f{\"u}r Astrophysik, \\
Karl-Schwarzschild-Str.\ 1, 85741 Garching, Germany}
\affiliation[3]{Section des Math\'ematiques, Universit\'e de Gen\`eve, 2-4 Rue du Li\`evre,  CH--1211 Gen\`eve 4, Switzerland}
\emailAdd{mona.frommert@unige.ch}

\date{\today}

\abstract{
We investigate the statistics of the cosmic microwave background using the 
Kolmogorov-Smirnov test. We show that, when we correctly de-correlate the data, the 
partition function of the Kolmogorov stochasticity parameter is compatible with
the Kolmogorov distribution and, contrary to previous claims, the CMB data 
are compatible with Gaussian
fluctuations with the correlation function given by standard $\La$CDM.
We then use the Kolmogorov-Smirnov test to derive upper bounds on
residual point source power in the CMB, and indicate the promise of this statistics
for further datasets, especially Planck, to search for deviations from Gaussianity and for detecting
point sources and Galactic foregrounds.
}


\maketitle
\flushbottom  

\section{Introduction}

The cosmic microwave background (CMB) is the most precise, best understood 
and therefore also most precious dataset in cosmology. Within the accuracy of
present observations, the CMB exhibits a 
perfect blackbody spectrum of radiation at the temperature 
$T_0=(2.72548 \pm 0.00057)$K~\cite{temp} 
with small, Gaussian fluctuations and a tiny amount of polarisation. The anisotropies and 
the polarisation can to a large extent be calculated within linear 
perturbation theory. Therefore, if the initial conditions are Gaussian as
predicted from simple inflationary models, the temperature 
anisotropies and the polarisation should represent Gaussian random fields
in the sky.

In this paper we outline a new method to test the Gaussianity of the CMB sky 
using the well-known Kolmogorov-Smirnov test. We describe how to de-correlate
Gaussian variables so that the Kolmogorov-Smirnov test can be applied. We also discuss in detail practical issues one has to deal with when applying the Kolmogorov-Smirnov test to CMB data. Such practical issues include finite-size effects,  numerical issues for highly correlated data, effects from determining the covariance matrix from the same dataset which is to be de-correlated, and correlations between different subsets of the CMB map for which we compute the Kolmogorov stochasticity parameter.
Not taking into account these effects can lead to wrong conclusions about the result of the Kolmogorov-Smirnov test.
Once all of these issues are taken into account properly, the failure of the Kolmogorov-Smirnov test 
would be an indication that the data is non-Gaussian and it can be used to quantify higher order correlations. 

When applying the Kolmogorov-Smirnov test to data from the Wilkinson Microwave Anisotropy Probe (WMAP), we find no indication of such a failure; the maps seem to be entirely compatible with Gaussian fluctuations whose correlation function is given by the standard $\La$CDM model.
Note, though, that since the significance of the outcome of the Kolmogorov-Smirnov test is limited by the resolution of the CMB map, this result might change with future CMB data as e.g.\ from the Planck satellite. 

We further use the Kolmogorov-Smirnov-test to obtain upper bounds on residual point source power in the CMB. The results are not yet competitive to the WMAP constraints, but they are derived for every band separately, without assuming a given scaling of the point-source power with frequency as in the WMAP papers. There are also ways of improving these constrains, which we plan to explore in future work.

In the next Section we introduce the  Kolmogorov-Smirnov test, discuss its 
modification for correlated variables, and study in detail different effects that can change the results of the test.
In Section~\ref{s:CMB}  we apply it to  simulated and measured CMB data.
In Section~\ref{s:ps}, we use the Kolmogorov-Smirnov test to obtain upper bounds on residual radio point sources in the CMB.
In Section~\ref{s:con} we conclude and discuss future applications of the method 
outlined here.

\section{The Kolmogorov-Smirnov test}\label{s:kol}

\subsection{Basics}

The Kolmogorov-Smirnov test is based on the following mathematical 
theorem~\cite{kolm}:\\
{\bf Theorem:} (Kolmogorov)\\
Let $\{X_1,\cdots, X_n\}$ be $n$ {\bf independent} realizations of a real random
variable $X$ ordered such that $X_1\le X_2 \le \cdots \le X_n$ and be
\be
F(x) = P\{X\le x\} \qquad (P=\mr{probability})
\ee
the partition function of $X$. The empirical partition function is given by
\be
F_n(x) := \left\{\begin{array}{ll}
         0 & \mbox{ if } x<X_1 \\
         k/n &\mbox{ if } X_k\le x< X_{k+1} \\
          1 &\mbox{ if } X_n\le x \ .
\end{array} \right.
\ee
The  Kolmogorov stochasticity parameter  is defined by
\be
\la_n :=\sqrt{n}\sup_x|F_n(x)-F(x)|\,.
\label{def_la}
\ee
 Its partition function, denoted by $\Phi(n,\la)$,
\be
 \Phi(n,\la) = P\{\la_n\le \la\}\,,
\ee
converges uniformly to the Kolmogorov distribution $\Phi_K(\la)$
given by
$\Phi_K(\la):=0$ for $\la\le 0$ and
\be
\Phi_K(\la) :=\sum_{k=-\infty}^{k=\infty}(-1)^k\exp(-2k^2\la^2) \quad 
\mbox{for } \la>0 \, .
\ee
\be
\lim_{n\ra\infty}\Phi(n,\la)=\Phi_K(\la)\,.
\ee
The Kolmogorov distribution, $\Phi_K$, is independent of $F$.

This theorem holds whenever the partition function $F$ is continuous.

We would like to apply the theorem to the temperature fluctuations 
$\De T(\bn)$ in different directions $\bn$, but there are two conditions of 
the theorem which are not satisfied a priori: 

First, the temperature fluctuations, even if 
they are a realization of the same stochastic process at each point (this 
is a consequence of stochastic isotropy),  are not independent at 
different points. The CMB sky is correlated, 
$$ \lan\De T(\bn_1)\De T(\bn_2)\ran = C(\theta)\neq 0 \, , \quad 
 \cos\theta=\bn_1\cdot\bn_2 \,. $$ 

Secondly, we do not know the theoretical distribution $F$ of CMB anisotropies. 
We assume it to be Gaussian with mean zero but the variance, $C(0)$ is 
determined from the data. Hence we compare the data with an empirically 
obtained distribution $F$. Clearly, the Gaussian distribution $F$ obtained 
by using the variance of the data will in general be closer to the empirical 
distribution $F_n$ than statistically expected. 

In addition, there are several practical issues such as finite-size effects and numerical issues for highly correlated data, 
which can change the results of the Kolmogorov-Smirnov test significantly. 
We discuss all of these effects in more detail in Sections~\ref{sec:effects} and \ref{ss:dec} and in the Appendix.

In the literature the Kolmogorov-Smirnov test has been applied to the CMB without
any discussion of the above-mentioned issues~\cite{Gur}. We  believe
that the conclusions drawn in these papers are not reliable because of that.
Recently the first point and the issue of finite-size effects have  been addressed in Ref.~\cite{Naess}.
We shall comment on this in more detail below.

\subsection{Effects of correlated variables, finite-size, and empirical determination of parameters}\label{sec:effects}

It is not difficult to guess what correlations will do to the Kolmogorov 
 stochasticity parameter: positive correlations will favor the variables 
$X_j$ to cluster. For a fixed number $n$ they thus explore less of 
the space of values than uncorrelated variables and
therefore $F_n$ is expected to deviate more from the theoretical distribution 
$F$ and the Kolmogorov parameter $\la_n$ will in general be larger. If
the correlations are negative the opposite effect is expected and the 
stochasticity parameter $\la_n$ is reduced.

\begin{figure}
\centering
\includegraphics[width=7.5cm]{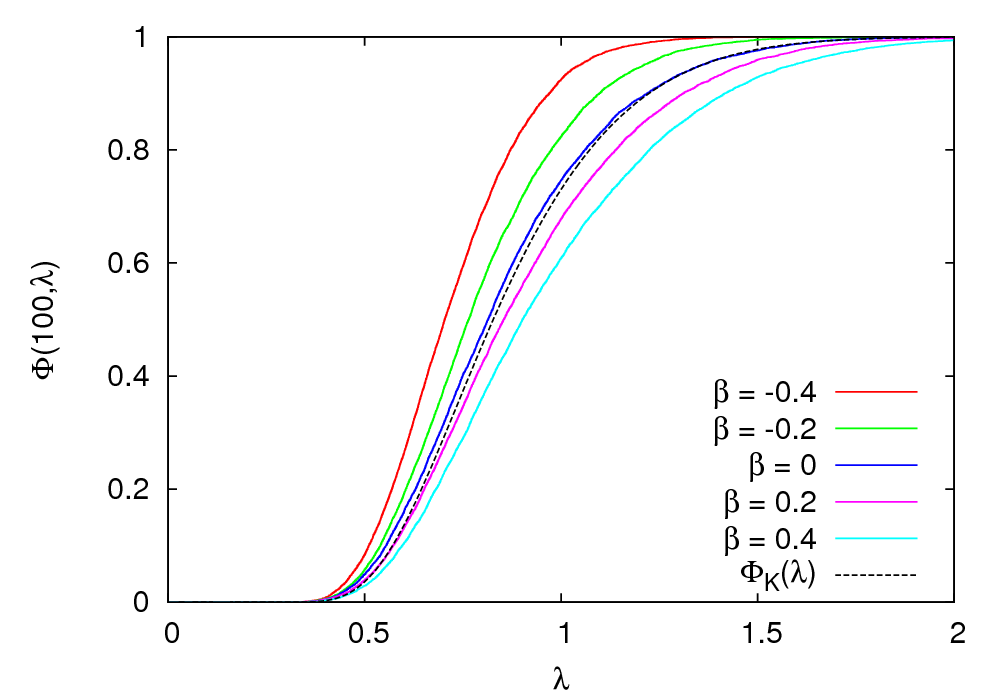}
\caption{\label{f:theta} The partition function for $\la_n$ as function of the
correlation parameter $\beta$ for $n=100$. Curves from left to right correspond to 
$\beta=-0.4,~-0.2,~0,~0.2,~0.4$. The curve for $\beta=0$ (middle, dark blue)
agrees with the Kolmogorov distribution $\Phi_K(\la)$ (dashed black line). We have used 10000 
realizations of the model to explore the partition function $\Phi(100,\la)$
for each value of $\beta$. As we have argued above, for positive correlations 
the distribution function is shifted to the right (to higher values of $\la_n$), 
while it is shifted to the left for negative correlations.
}
\end{figure}

This result is illustrated in Fig.~\ref{f:theta}, where the partition function
of $\la_n =\sqrt{n}\sup_y|F_n(y)-F(y)|$ is show for $n=100$ for a model with
nearest neighbor correlations,
\be
 \lan Y_jY_i\ran =\left\{ \begin{array}{ll} 
    0 & \mbox{ if }\quad |i-j|>1 \\
    \beta  & \mbox{ if }\quad |i-j|=1 \\ 
      1 & \mbox{ if }\quad  i=j \, . \end{array} \right.
\ee
We draw independent random variables $X_j$ from a Gaussian normal distribution with vanishing mean and
variance one, $N(0,1)$. The correlated variables $Y_j$ are obtained by following
the opposite of the procedure outlined in Section~\ref{ss:dec} below on 
uncorrelated variables. Another, analytically solvable model is presented 
in Ref.~\cite{master}. Note that throughout this paper, we will denote correlated random variables by $Y_j$, whereas $X_j$ are understood to be independent variables drawn from $N(0,1)$.

\begin{figure}[h]
\centering
\includegraphics[width=7.5cm]{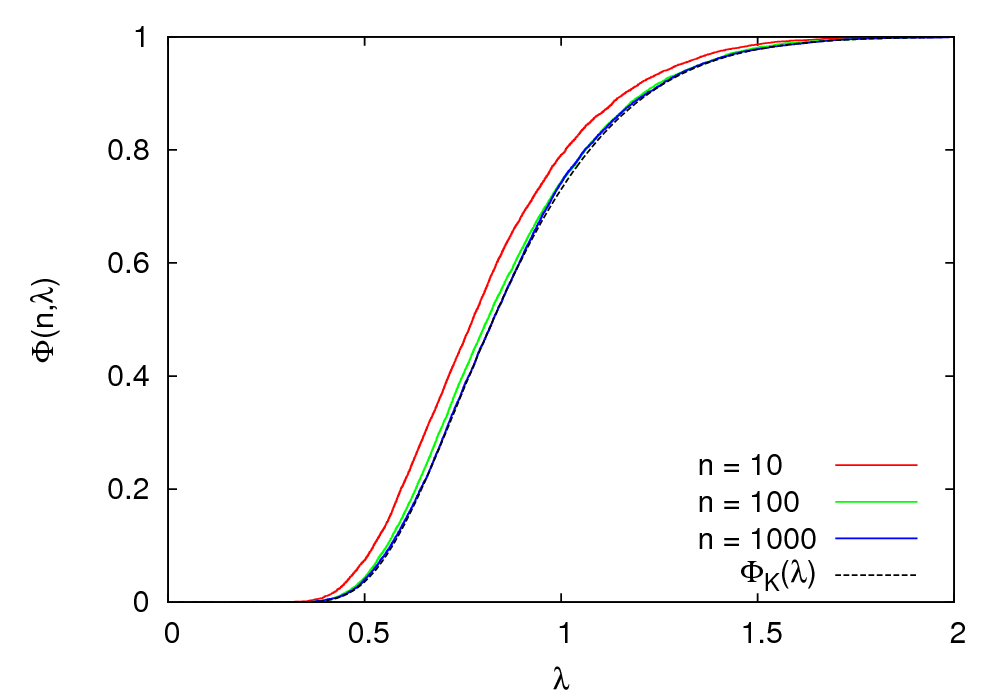}
\includegraphics[width=7.5cm]{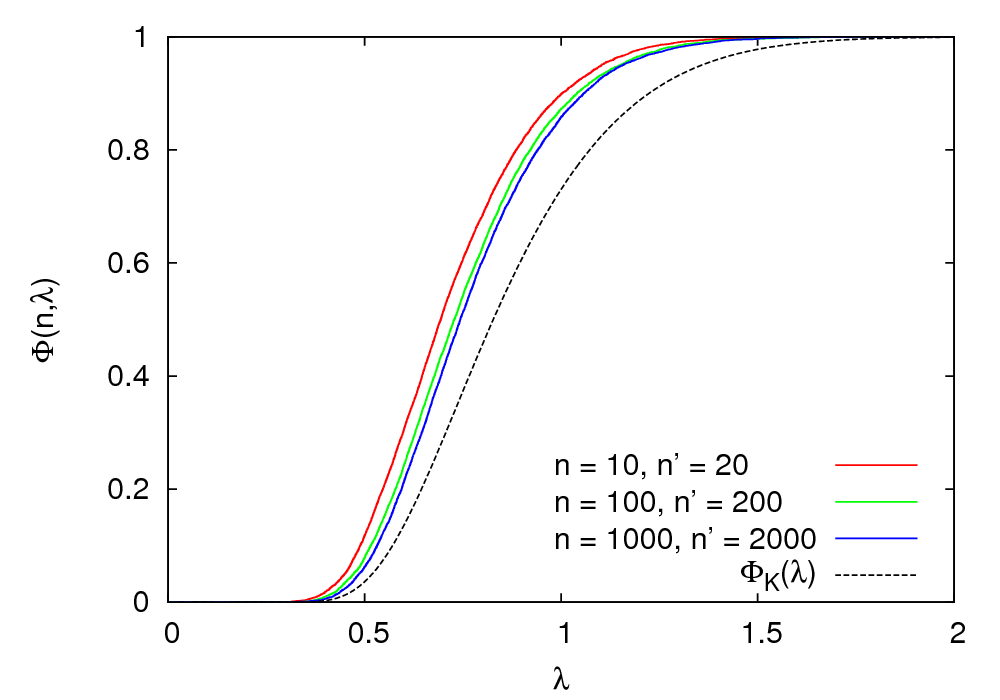}
\includegraphics[width=7.5cm]{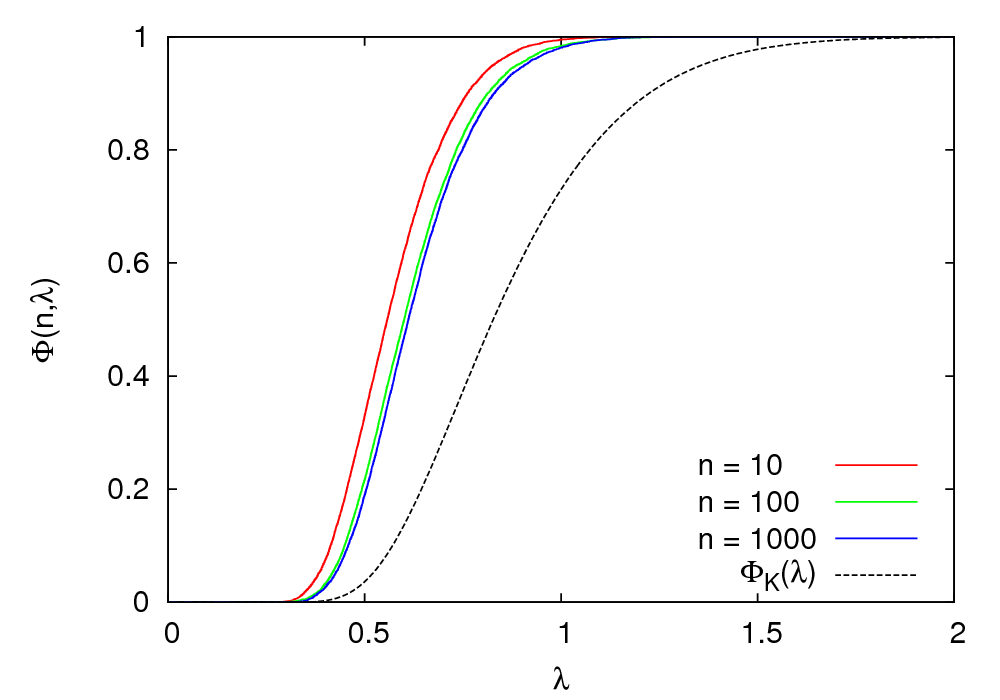}
\caption{\emph{Upper left:} Partition function $\Phi(n,\la)$ 
for $n$ independent random varibales $X$ for $n=10$, 100, and 1000. Superimposed 
is the Kolmogorov function $\Phi_K(\la)$. \newline
\emph{Upper right:} $\Phi(n,\la)$ for $n$ independent random variables
if the partition function $F$ is taken to be $F=N(a,b)$ where
the mean $a$ and variance $b$ are determined from the data. The number 
of $X_i$ used to determine $a$ and $b$ is twice the one used 
to compute $\la_n$, $n' = 2n$. \newline
\emph{Bottom:} Same as in the middle panel, but with $n' = n$.}
\label{f:nocor} 
\end{figure}

To compare with the correlated results we also show the partition function 
$\Phi(n,\la)$ for the uncorrelated variables $X$ in the upper left panel of Fig.~\ref{f:nocor}. This is 
useful in order to separate effects from correlations from those of finite 
size (i.e.\ small $n$). For $n=100$, the finite-size effects are already very small (the curve is already very close to the Kolmogorov-curve). However, if we compute the mean of $\Phi_K(\la_n)$ for $n=100$ from $m=10000$ samples, it is $\langle \Phi_K \rangle \approx 0.48$ rather than 0.5, as we would expect for the limiting case of $n \rightarrow \infty$. This is only a difference of a few percent, however, this difference will become relevant for us later on. We study the effects of small $n$ in much more detail in Appendix \ref{sec:finite_size}.

Intuitively it is also clear what happens if we determine the parameters of the theoretical
distribution function from the data which we also use to calculate $F_n$:
on average the distance $\la_n$ between $F_n$ and $F$ will be smaller
and therefore also $\Phi_K(\la_n)$ is smaller, see also \cite{liliefors}. 
In the upper right and bottom panels of Fig.~\ref{f:nocor}, we show the  partition function $\Phi(n,\la)$ obtained if $F$ is not taken to be $N(0,1)$ but $N(a,b)$ where we determine the mean $a$ and the variance $b$ 
from the data,
\be
 a = \frac{1}{n'}\sum_{i=1}^{n'} X(\bn_i) \,, \qquad
 b =  \frac{1}{n'-1}\sum_{i=1}^{n'} (X(\bn_i) -a)^2\,.
\ee
In this case, $F_n$ is by construction too close to $F$, thus moving $\Phi(n,\la)$ to the left. This effect is
quite strong if we determine the parameters from the same set of $X_i$ that we use to compute $\la_n$ (c.f.\ 
bottom panel of Fig.~\ref{f:nocor}, where we used $n' = n$) and becomes less pronounced if we use additional $X_i$ to 
compute the mean and the variance (upper right panel, $n' = 2n$).

We suspect that such an effect is the 
reason why the authors of Refs~\cite{Gur} have obtained so small values 
for $\Phi_K(\la_n)$ especially for relatively small $n$, see also~\cite{Naess}.
With increasing $n$  the value $\Phi_K(\la_n)$ calculated in Refs.~\cite{Gur} 
grows, approaching  unity in two cold spots with $5^\circ$ radius.
We suggest that this is due to the correlations of the temperature 
anisotropies in this region. In the next Section we shall show this
quantitatively.

\subsection{De-correlating Gaussian variables}\label{ss:dec}

Let us consider an $n$-dimensional Gaussian variable $Y$ with vanishing mean
but with correlations. Its probability density is then given by
\be\label{e:corr}
f_Y(y) =\frac{1}{\sqrt{(2\pi)^n\det C}}\exp\left(-\frac{1}{2}y\cdot 
 C^{-1}y\right)
\, , \quad y\in \mathbb{R}^n \, ,
\ee
where $C$ is the symmetric, positive definite correlation matrix,
\be
 \lan Y_iY_j\ran = \int\! d^ny \, y_iy_j \, f_Y(y) = C_{ij} \,.
\ee
We can diagonalize this matrix with some orthogonal matrix $O$ so that
\be\label{e:diag}
C = OMO^T \quad M =\left(\begin{array}{ccc}
                              \mu^2_1 &\cdots& 0\\
                              0 &\ddots &0  \\
                              0 &\cdots &\mu^2_n \end{array}\right) \, .
\ee
If $Y$ is Gaussian, the $n$ variables $X_i$ defined by
\be\label{e:decor}
X_i = \frac{1}{\mu_i}(O^TY)_i
\ee 
are uncorrelated and distributed with the normal distribution $N(0,1)$. 
On these variables the Kolmogorov-Smirnov 
test should apply, i.e., the partition function of the stochasticity
parameter $\la_n$ should converge to $\Phi_K$ for large $n$. 

If the partition function $\Phi(n,\la)$ for $\la_n$ obtained by comparing 
the empirical distribution of $X$ with  $N(0,1)$  does not converge to the 
Kolmogorov function $\Phi_K$ there are two possibilities.
\begin{itemize}
\item We have not used the correct correlation matrix $C$ (which is typically
determined by the cosmological parameters).
\item The fluctuations are not Gaussian, i.e.\ there are also higher order
correlations such that even though $\lan X_iX_j\ran =\de_{ij}$, Wick's theorem 
is not satisfied for higher order correlators.
\end{itemize}
In either case this would falsify the hypothesis that CMB fluctuations are
well explained by Gaussian inflationary perturbations in a $\La$CDM model 
with the cosmological parameters inferred in the literature.

We must, however, note that even though we do determine the correlation 
function from the cosmological parameters and not directly from the data,
these parameters are inferred using the same data and are therefore not
independent of it. Also in this approach, we are therefore somewhat subject to
the problem that we determine the distribution of our variables from the data
itself. In Appendix \ref{sec:lilliefors} we show that for the case of the 
CMB we do not see this effect. 
The reason is that the $\La$CDM model puts a strong prior on the form of the correlation function, leaving only six parameters to be determined from a large amount of available CMB data.

Note also that the de-correlation transformation, eq.~(\ref{e:decor}), has its limitations for very strongly correlated random variables. In the limiting case of two perfectly correlated random variables, $Y_1 = Y_2$, the covariance matrix $C$ has one zero eigenvalue, so that we cannot compute $X_i$ for this eigenvalue. In such a case, one needs to directly work with the correlated variables $Y$, and compare the partition function of their Kolmogorov parameter to simulated ones. This is what we call `correlation method' in the following.\footnote{An analytical investigation of such models with perfect correlations can be found in \cite{master}.} In cases where we do not have perfectly correlated data, but nevertheless very strong correlations (e.g.\ between neighbouring pixels), we can still run into numerical problems with the transformation in eq.~(\ref{e:decor}), since we divide by a very small number $\mu_i$. We investigate this effect for CMB maps with negligible noise in Appendix \ref{sec:pixelisation}: The empirical partition function of the Kolmogorov stochasticity parameter for the de-correlated variables can be significantly shifted to the right
due to very strong correlations.


\section{Application to the CMB}\label{s:CMB} 
In this Section we apply the Kolmogorov-Smirnov test to simulated data and to the seven year
CMB anisotropy data from WMAP.

\subsection{Simulated data}\label{sec:mock}

\begin{figure}
\centering
\includegraphics[width=7.3cm]{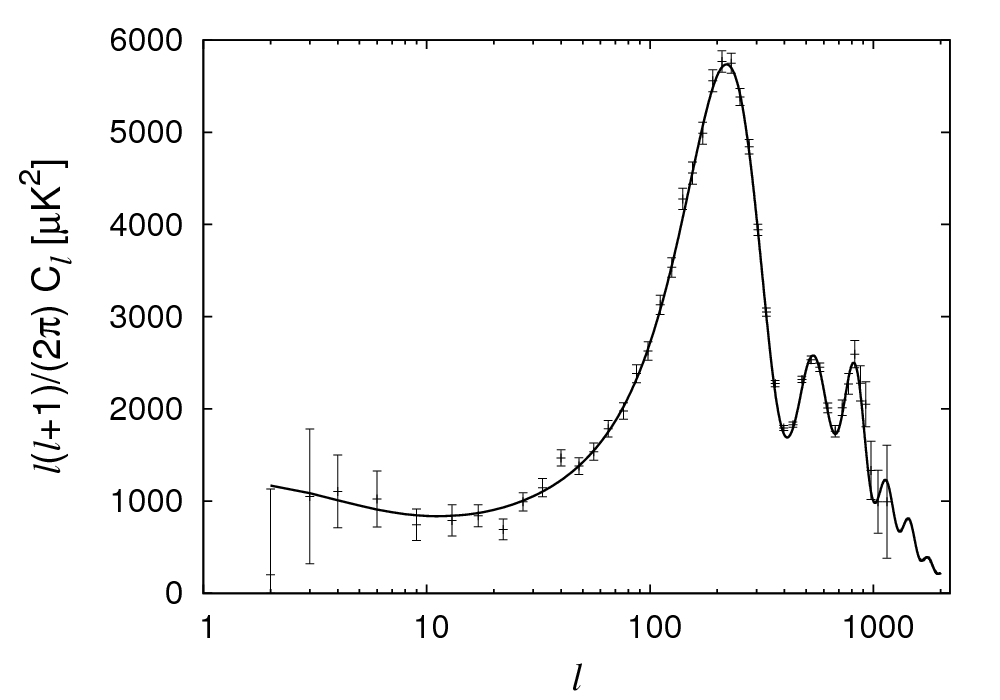} \quad
\includegraphics[width=7.3cm]{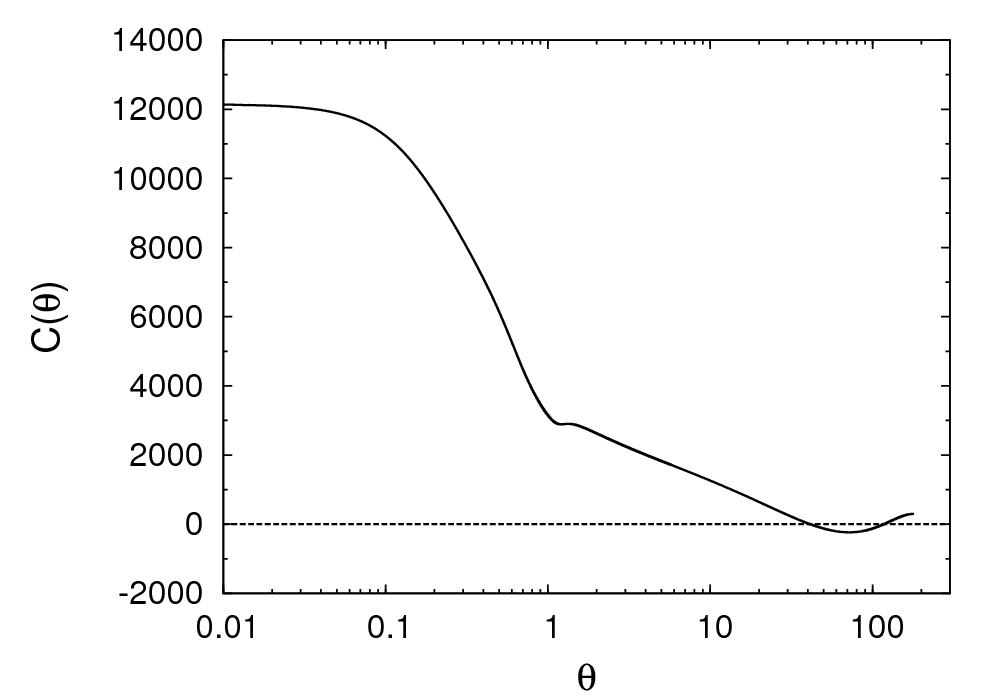}
\caption{\label{f:Cls} The $C_\ell$'s (left) and the CMB correlation function (right) from the WMAP seven year best fit 
cosmological parameters given above and the WMAP data with error bars,  $(\mu K)^2$. The `kink' 
in the correlation function at $\theta=1^\circ$ corresponds to the acoustic peaks in the power spectrum.}
\end{figure}

We start by calculating the CMB correlation function using the best fit
cosmological parameters of WMAP, as they are given in~\cite{Larson2011}, see 
 Table~\ref{t:parms}. 
\begin{table}[ht]
\centering
\begin{tabular}{cccl}
$n_s$ & $=$ &  $0.96$ & {scalar spectral index}\\
$\De_R $ & $=$ &  $2.4\times 10^{-9}$  & amplitude of the curvature\\
         &     &                      & perturbation spectrum\\
$r $ & $=$ &  $0$ & tensor to scalar ratio\\
$h^2\Om_b$ & $=$ &  $0.0226$ & baryon density\\
$h^2\Om_c $ & $=$ &  $0.111$ & cold dark matter density\\
$\tau $ & $=$ &  $ 0.088$ & optical depth to the \\
        &     &           & last scattering surface\\
$\Om_\La$ &=  &0.73 & cosmological constant\\
$\Om_k$ &=& 0.00 & curvature parameter
\end{tabular}
\caption{\label{t:parms}The (minimal) cosmological parameters from the WMAP
seven year data \protect\cite{Larson2011}.  } 
\end{table}
The last digit given is rounded and is uncertain. More details about
the error bars and the definition of these parameters can be found in 
Ref.~\cite{Larson2011}.
In Fig.~\ref{f:Cls} we compare the CMB angular power spectrum obtained with 
these parameters with the WMAP seven year data. A definition of the CMB angular 
power spectrum can be found e.g.\ in Ref.~\cite{mybook}

The correlation function for the temperature anisotropy of two directions 
$\bn_1$ and $\bn_2$ in the sky can be calculated from the CMB power spectrum,
see~\cite{mybook},
\be\label{e:corT}
\lan \De T(\bn_1)\De T(\bn_2)\ran = 
\frac{1}{4\pi}\sum_{\ell=2}^\infty (2\ell +1)C_\ell P_\ell(\bn_1\cdot\bn_2) \,,
\ee
where $P_\ell$ denotes the Legendre polynomial of degree $\ell$. The 
correlation function obtained from the spectrum of Fig.~\ref{f:Cls} is shown 
on the right hand panel.

We now simulate datasets consisting of $n$ CMB-`pixels':
We draw $n$ independent realizations $X(\bn_i)$ from the normal distribution $N(0,1)$,
which we then correlate with the help of the correlation matrix
\be
C_{ij} = C(\theta_{ij}) \,,
\ee
following the opposite of the procedure outlined in Section~\ref{ss:dec}.
Here, $\theta_{ij}$ is the angular separation of the points $\bn_i$ and $\bn_j$.
For the sake of simplicity, we assume all  of our data-points to lie on a great arc of opening 
angle $\al$.\footnote{
Note that working with data lying on a great arc is especially useful when we consider the full circle:
To transform the uncorrelated variables $X$ to the correlated ones given by
$Y_i =\mu_i(OX)_i$ where $O$ and $\mu_i$ are defined as in Eq.~(\ref{e:diag}),
we have to diagonalize the correlation matrix $C$. For large values of $n$ this
requires a significant numerical investment:
Diagonalization of a 
symmetric $n\times n$ matrix determining both, the eigenvalues $\mu_i^2$ 
and the orthogonal matrix $O$ requires $n^3$ operations.  
To sample the 
variable $Y$ and therefore $\la_n$ $m$ times requires ${\cal O}(m\cdot n^3)$ 
operations. If however the points $\bn_i$ are
equally spaced on a circle such that the angular separation of $\bn_1$ and 
$\bn_n$ is equal to the resolution $\de$, the matrix C becomes `circulant' and
its diagonalization corresponds to a Fourier transform. In this case the
number of operations can be reduced to  ${\cal O}(m\cdot n\log(n))$ which is 
of course a significant improvement. For more details see~\cite{master}.
}
For a given
number $n$ of points, nearest neighbors then have an angular separation
 $\de = \al/n$ and we have to include at least $\ell_{\max}>n\pi/\al$ 
$~C_\ell$'s in Eq.~(\ref{e:corT}) in order to be able to resolve this scale. In practice, we use $\ell_{\max} = 2000$ for the plots in this Section.\footnote{For $n=250$ and $n=500$ and an opening angle of $\alpha = 10^\circ$, $\ell_{\max} = 2000$  is not enough to completely resolve these scales; however, the result that a higher resolution of the map yields larger values of $\la_n$ is not affected by this.}$^,$\footnote{In all of the following sections, we include the beam of the WMAP-experiment, which cuts off power at $\ell \sim 500/700/900$ for the Q/V/W-band, respectively.}
For these CMB datasets, we obtain the partition function $\Phi(n,\la)$ of $\la_n$ from 10000 simulated datasets of $n$ `pixels' each.

\begin{figure}
\centering
\includegraphics[width=7.5cm]{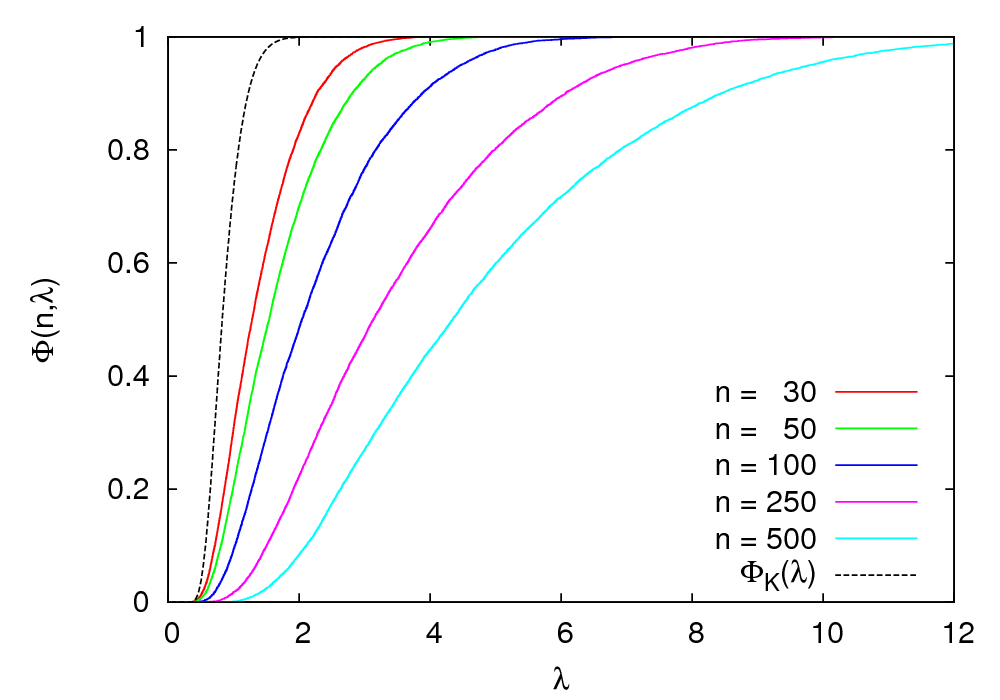}  \includegraphics[width=7.5cm]{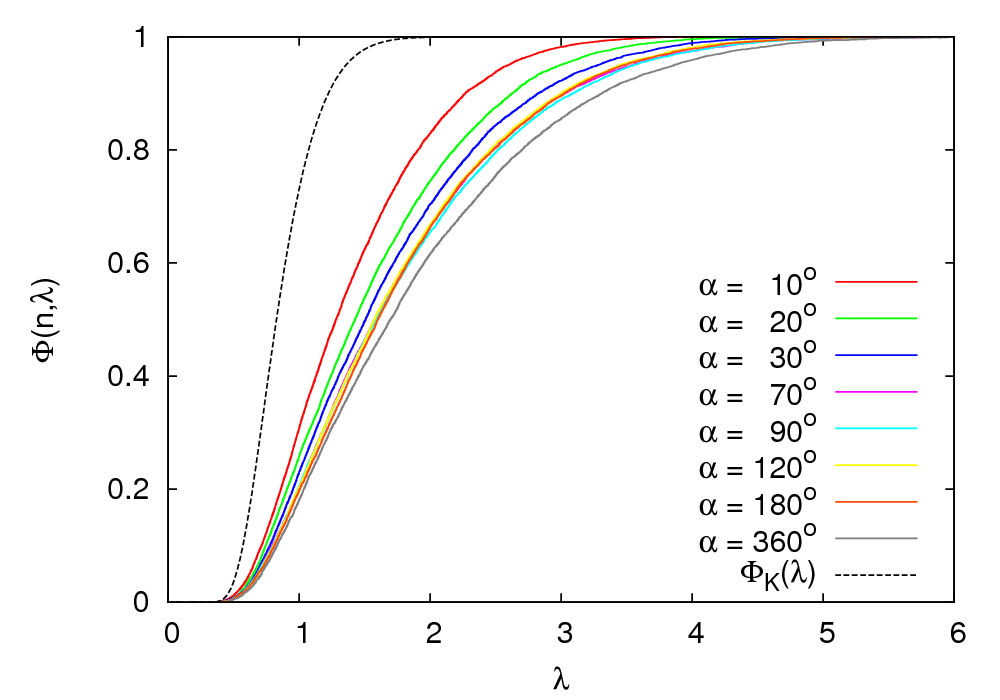}
\caption{\label{f:reso} \emph{Left:} The partition functions $\Phi(n,\la)$ for $n$ points
of simulated data in a total opening angle of $\al=10^\circ$ are shown for $n=30-500$. 
For higher $n$ the points are closer and therefore more correlated. Hence with 
increasing $n$,  $\Phi(n,\la)$ deviates more from the Kolmogorov function
$\Phi_K(\la)$. As explained in the text, positive correlations prefer larger
values of $\la$.\newline
\emph{Right: }The partition functions $\Phi(n,\la)$ for $n$ points
of simulated data at fixed resolution $\de=\al/n=20$ arc minutes. 
Until an opening angle of $\al=70^\circ$, which corresponds to $n=210$, 
the correlations increase with $n$ and 
$\Phi(n,\la)$ moves to the right, preferring larger values of $\la$. For 
larger angular separation the anti-correlations present for angular separations 
$50^\circ \lesssim \theta \lesssim 100^\circ$ come in and balance the correlations, such that 
$\Phi(n,\la)$ is kept at roughly the same position up to an opening angle of $\al=180^\circ$.
For $\al=360^\circ$, corresponding to $n=1080$, the correlations have moved $\Phi(n,\la)$ 
further to the right again.\newline
In both plots the Kolmogorov function $\Phi_K$ is also indicated (dashed black line).
}
\end{figure}

On the left plot in Fig.~\ref{f:reso} we show the partition functions $\Phi(n,\la)$ of
the temperature fluctuations for points on a great arc with fixed total opening 
angle $\al=10^\circ$ for different values of $n$. As the number of points, $n$, 
increases, the angular distance between nearest neighbors, $\de=\al/n$, decreases
and they are more and more correlated, see Fig.~\ref{f:Cls}. As we have 
explained in Section~\ref{s:kol}, this means that the probability for higher 
values of $\la_n$ becomes larger and hence the function $\Phi$ is shifted to the 
right with respect to the Kolmogorov function $\Phi_K$ which is also indicated 
as the dashed line in Fig.~\ref{f:reso}. Therefore, if in a fixed patch the number 
of points is increased, the typical 
value of $\la_n$ increases and with it $\Phi_K(\la_n)$. At this level, this
has nothing to do with non-Gaussian features of the CMB but is simply an 
effect of correlations. 

On the right plot in Fig.~\ref{f:reso} we fix the resolution to $\de=20$ arc minutes and simply 
increase the number of points. In this case, the variation of $\Phi(n,\la)$
with $n$ is much less pronounced. Here it comes from the fact that the total 
opening angle $\al =n\de$ depends on $n$. 
For $n\le 210$, which corresponds to $\al\le 70^\circ$, 
the function $\Phi$ is still moving to the right with 
increasing $n$, i.e., correlations are becoming stronger; while for 
$210<n<540$, i.e.\ for $70^\circ<\alpha<180^\circ$  
the small anti-correlations which are present for angles 
$50^\circ\lesssim\theta\lesssim 100^\circ$, see Fig.~\ref{f:Cls}, balance the correlations such that the curves $\Phi$ stay at the same position, or even move slightly to the left. 
For $n=1080$, corresponding to $\al=360^\circ$, the correlations have moved $\Phi$ further to the
right again. Note that the partition functions for $\alpha = 10^\circ$ and $20^\circ$, corresponding to $n=30$ and $n=60$ are still subject to effects of small $n$, which shifts their partition functions to the left.

 
\subsection{WMAP - local analysis}\label{ss:WMAP}

\subsubsection{General remarks}

In the previous Section we have studied the partition function $\Phi(n,\la)$ for ring segments of a given opening angle of a simulated CMB map. Here we want to locally analyze the observed data. 
In the following we denote a $\la_n$ computed locally for a given patch or ring segment consisting 
of $n$ pixels by $\la_n^\loc$, as opposed to the global $\la_n$, which we introduce 
in Section~\ref{s:global}. 
$\la_n^\loc$ could turn out to be  useful for detecting local non-Gaussianities such as Galactic foregrounds, see also \cite{Gur, Naess}.

We  have two possibilities to apply the Kolmogorov-Smirnov test to the WMAP data: We can either de-correlate the pixels in 
a given patch or ring segment using Eq.~(\ref{e:decor}) and compare the 
$\la_n$ to the Kolmogorov distribution $\Phi_K(\la)$. Or we can directly work with the (correlated) 
temperature values in the CMB map and compare to our simulated partition functions $\Phi(n,\la)$
 of a correlated signal plus noise. 
In the following, we will present both methods. For the `correlation method', we will work with ring segments as we did for the simulations described in Section \ref{sec:mock}, because this facilitates the simulation of the partition function. For the `de-correlation method', we will use patches because they are better localized than the ring segments, thus being more suitable for a local study of the CMB. However, in Appendix \ref{app:decorr}, we will also perform the de-correlation method on ring segments, in order to compare the results to the ones obtained from the correlation method.

In general, we consider the `de-correlation method' to be preferable over the `correlation method'. First of all, in this case
there is no need to simulate the partition function of $\la_n$ since it is the well-known Kolmogorov distribution.
The de-correlation  method is also generally more powerful than the correlation method, since it
is less degenerate in the detection of the different effects that can cause a deviation from the
Kolmogorov curve. Consider, for example, a dataset for which the empirical partition function of the $\la_n$'s lies to the 
right of the Kolmogorov curve $\Phi_K(\la)$. This could be due to correlations or to other effects such as non-
Gaussianity or to the fact that we have used a wrong $F(x)$ when computing the $\la_n$'s. In simulations of correlated 
data, we can probably mimic this effect by using a certain correlation function. However, if the shift of the curves is not 
due to correlations in the data, we would in general not expect to obtain the Kolmogorov distribution when de-correlating the data with any correlation function.
Furthermore, with the de-correlation method it is straightforward to account for 
inhomogeneous noise.

However, note that the correlation method is more stable in the case of very strongly correlated pixels. Pixels can be very strongly correlated if the noise in the CMB map is negligible. In that case, the de-correlation method can lead to highly biased results, as we show in Appendix \ref{sec:pixelisation}. If the eigenvalues of the covariance matrix are extremely close to zero, it might even not work at all. For WMAP, we do not encounter this problem, because we have enough noise power on small scales. However, for future CMB experiments, with less noise, this issue might become relevant.

For both methods, we compare the results obtained from WMAP data to simulations of 100 CMB maps plus Gaussian 
white noise (mimicking the detector noise). For the de-correlation method, we use the correct (inhomogeneous) noise 
variance of the maps, whereas for the correlation method, the noise variance has been
averaged over all pixels used in the analysis.
In the correlation method, adding uncorrelated noise to the simulations of correlated data reduces the correlation of the 
data and thus shifts the partition function $\Phi(n,\la)$ to the left as compared to simulations without noise. As for the de-correlation method, the inclusion of noise just slightly broadens the scatter of the curves without introducing a bias.

We apply the Kolmogorov-Smirnov test to the foreground-cleaned seven year WMAP Q-band (40.7 GHz), V-band (61 GHz), and W-band (94 GHz) maps.

\subsubsection{Beam and detector noise}

When computing the correlation function from the CMB power spectrum, we need to take into account the beam transfer 
function of WMAP. This cuts off power at high $\ell$, and it significantly reduces the power even in the first peak, see 
e.g.~\cite{shanks}. 
Note that the correlation function shown in Fig.~\ref{f:Cls} is computed without taking into account the WMAP beam. 

Finally, we need to take into account the detector noise of WMAP.
If we assume that the maps have been perfectly cleaned from Galactic foregrounds and point sources, the data (i.e.\ the pixels of a given ring segment) are given by
\be
Y_i = \Delta T_i + n_i \,,
\ee
where $\Delta T_i$ and $n_i$ denote the CMB temperature and the detector noise in pixel $i$, respectively. If we ignore small non-Gaussianities coming e.g.\ from inflation, the data are Gaussian distributed, with the covariance matrix
\be
C_{ij} = \lan Y_i Y_j \ran = \lan \De T_i\De T_j\ran + N_{ij} \,,
\ee
where $\lan \De T_i\De T_j\ran$ is the CMB correlation function shown in Fig.~\ref{f:Cls}, and $N$ is the diagonal noise covariance matrix
\be
N_{ij} = \lan n_i n_j \ran = \si_i^2 \, \delta_{ij}  \,.
\ee
Here, $\si_i^2 = \si_0^2/N_{\rm obs}(i)$ is the detector noise variance in pixel $i$, which is inversely 
proportional to the number of observations of a given pixel, $N_{\rm obs}(i)$ \footnote{$\si_0$ for 
every frequency band can be found at \footnotesize
{http://lambda.gsfc.nasa.gov/product/map/dr4/skymap\_info.cfm}}.


\subsubsection{Results: de-correlation method}\label{s:decorr}

\begin{figure}
\centering
\includegraphics[width=7.5cm]{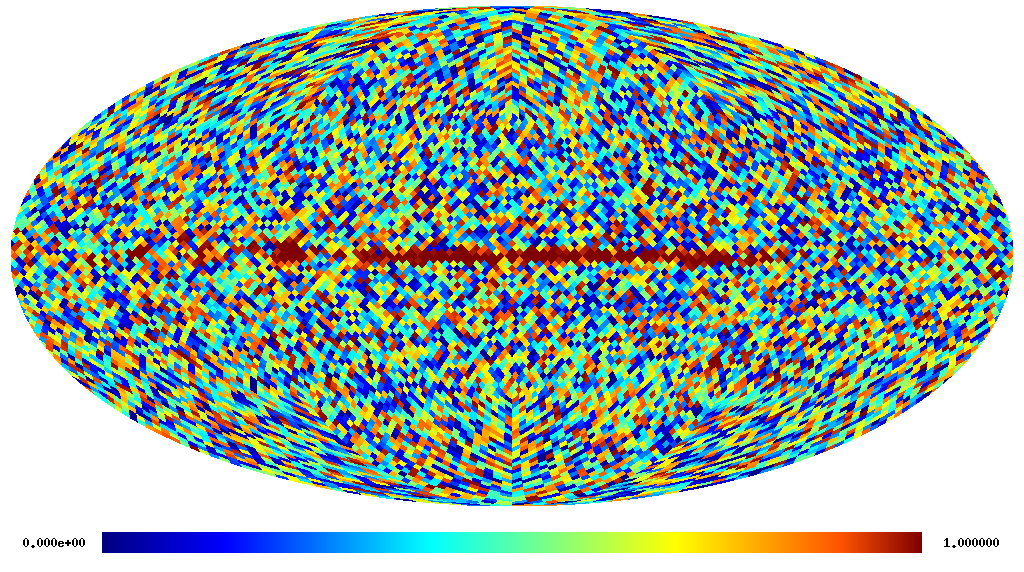}
\includegraphics[width=7.5cm]{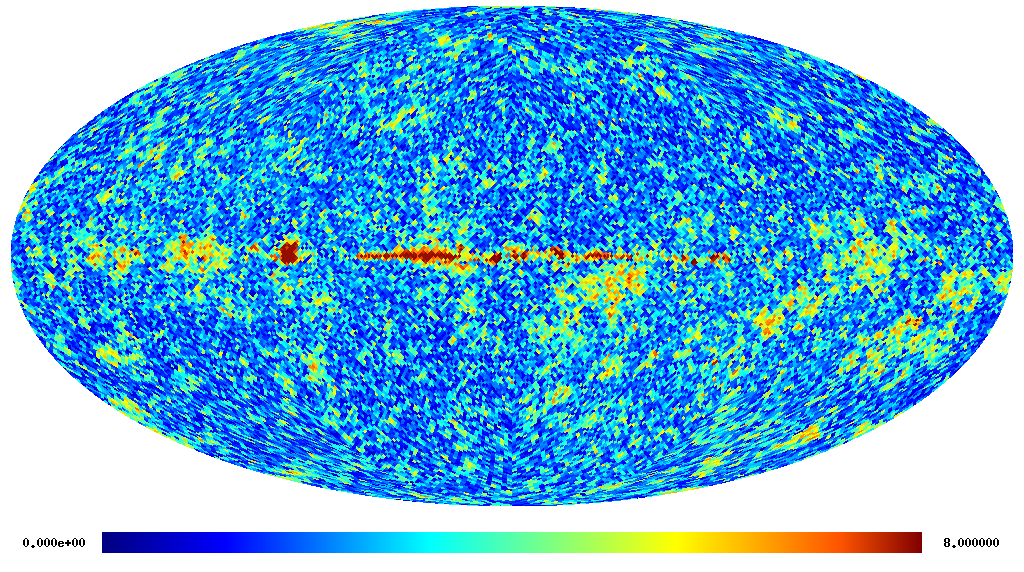}
\caption{\emph{Left:} Map of $\Phi_K(\la_n^\loc)$ for the V-band for the de-correlation method. The color scale goes from 0 to 1. As expected, outside of the galaxy, this map is very uniform. \newline
\emph{Right:} Map of $\la_n^\loc$ for the V-band for the correlation method. The color scale goes from 0 to 8.
In this map structure not due to the galaxy but due to correlations is clearly visible.}
\label{f:kolmap}
\end{figure}

\begin{figure}
\centering
\includegraphics[width=7.5cm]{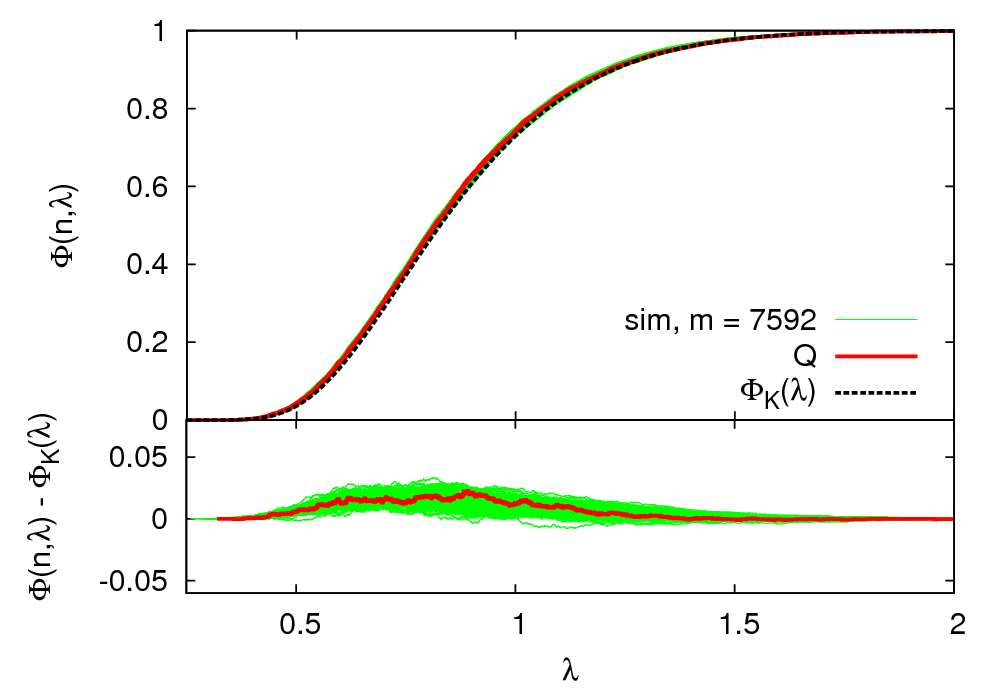}
\includegraphics[width=7.5cm]{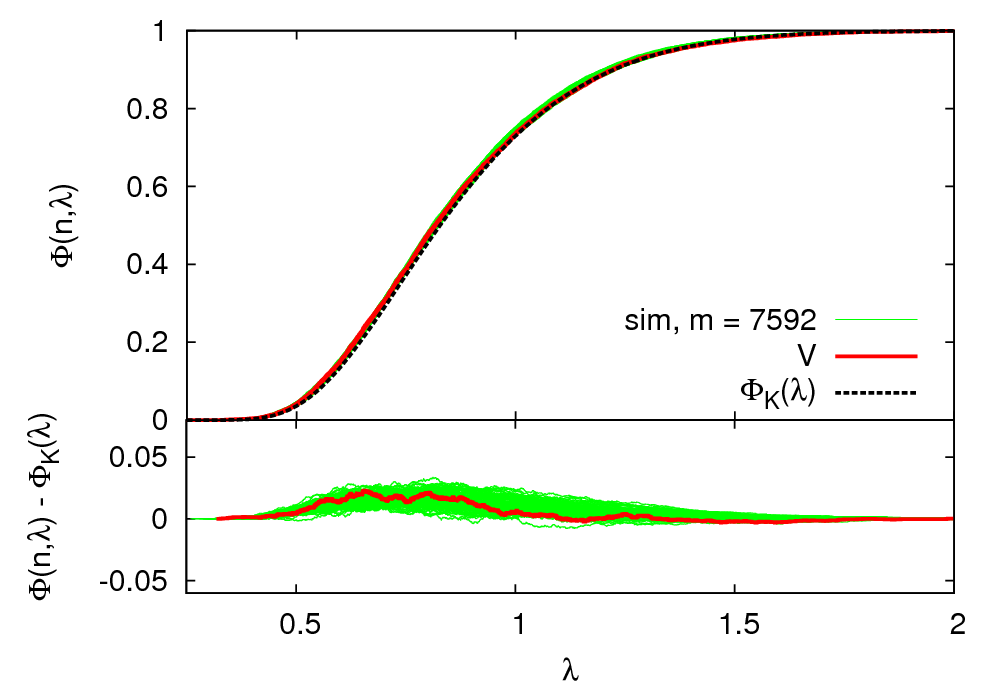}
\includegraphics[width=7.5cm]{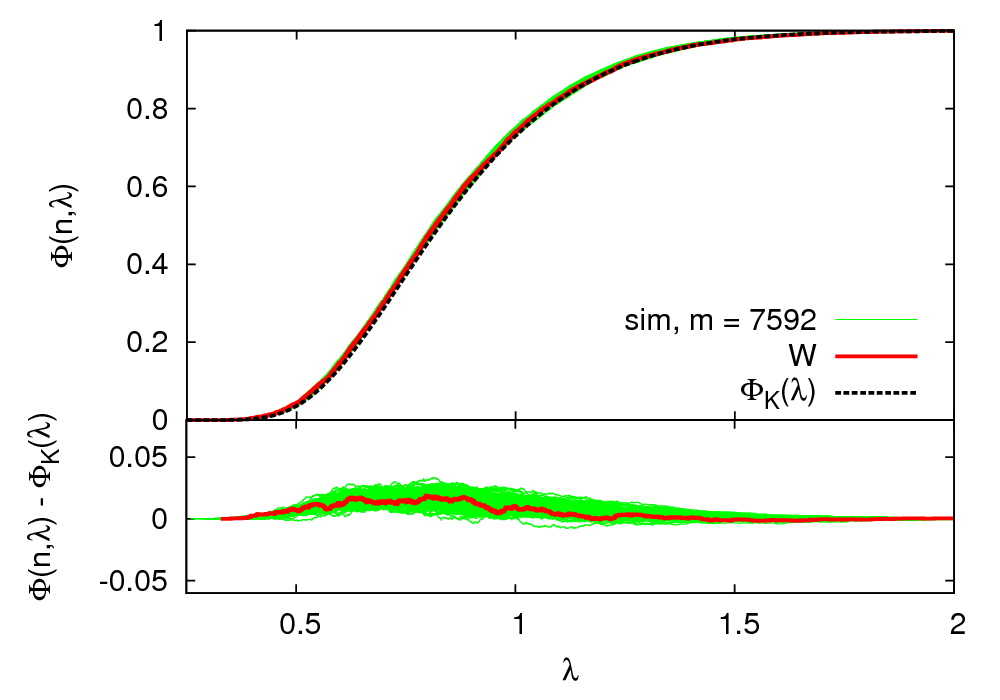}
\includegraphics[width=7.5cm]{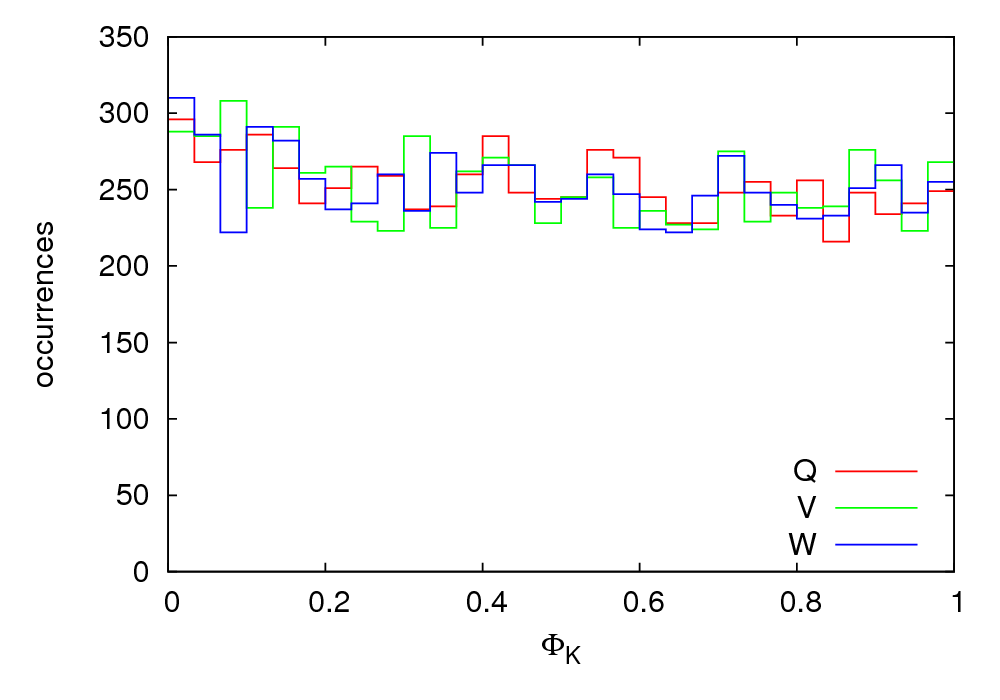}
\caption{De-correlation method: empirical partition function of $\la_n^\loc$ for WMAP-7year-data and 100 simulations for the Q-band (\emph{upper left}), the V-band (\emph{upper right}), and the W-band (\emph{bottom left}).
 We have $m=7592$ patches and $n=256$ pixels per patch if none of the pixels is masked. We mask pixels using the WMAP extended mask; a patch is discarded if more than ten per cent of its pixels are masked.
 The simulations are shown in green. \newline
\emph{Bottom right:} histogram of $\Phi_K$ for all bands. We see a small bias towards small values of $\la$
(larger $\Phi(n,\la)$ at fixed $\la$), reflecting the effects of small $n$.}
\label{f:WMAPdecorPatch}
\end{figure}

\begin{table}
\centering
\begin{tabular}{|c| c |  c  c  c  |}
\hline
method &  & Q & V & W \\
\hline
\hline
local (patches) & mean[$\Phi_K(\la_n^\loc$)] & 0.488 & 0.489 & 0.490 \\
\hline
 & stddev[$\Phi_K(\la_n^\loc)$] & 0.290 & 0.293 & 0.292 \\
\hline
\hline
global & mean[$\Phi_K(\la_n$)] & 0.508 & 0.505 & 0.500 \\
\hline
 & stddev[$\Phi_K(\la_n)$] & 0.291 & 0.284 & 0.284 \\
\hline
\end{tabular}
\caption{\label{t:la} Empirical mean and standard deviation of $\Phi_K(\la_n)$.
\emph{Top:} local analysis (de-correlation method), $n\approx 256$, $m=7592$.
\emph{Bottom:} global analysis, $n=2000$, $m=500$.
The empirical mean should be Gaussian distributed around $\langle  \Phi_K \rangle = 0.489$ ($0.495$) with a standard deviation of $\sigma \approx 0.0033$ ($0.013$) for the local analysis (the global analysis).
The empirical standard deviation should be about $\sqrt{\langle ( \Phi_K - \langle  \Phi_K \rangle)^2 \rangle} = 1/\sqrt{12} \approx 0.29$ for both. Both analyses are fully consistent with theoretical expectations.} 
\end{table}

For analyzing the WMAP data locally using the de-correlation method, we choose our patches to be the pixels of a Healpix-map with NSIDE=32, which gives us 256 pixels per patch. We discard all pixels masked by the WMAP extended mask, which masks also identified point sources outside our Galaxy, and we discard a patch if more than 10 per cent of the pixels in a patch are masked. This leaves us with $m=7592$ patches.
We de-correlate the data from every patch using Eq.\ (\ref{e:decor}), which should give us independent Gaussian random variables $X_i \sim N(0,1)$. Then, we compute the local $\la_n^\loc$ for every patch. 
The obtained map of $\Phi_K(\la_n^\loc)$ for the V-band (without applying the mask) is shown in Fig.~\ref{f:kolmap}. Our Galaxy is clearly visible, but outside the Galaxy it looks like white noise, as it should. The maps for the other bands look very similar. In the following, we apply the mask as explained above.

In order to get a visual impression of the distribution of the $\la_n^\loc$ of the different patches, we show their 
empirical partition function together with the Kolmogorov-curve $\Phi_K(\la)$ in Fig.~\ref{f:WMAPdecorPatch}. Also shown 
are the analysis from 100 simulated CMB maps to which Gaussian white noise has been added. For the simulations we consider homogeneous noise, whereas we account for 
inhomogeneous noise when de-correlating the real data. All three bands of real data lie well within the expected 
variance. Note that the simulated curves are biased to the left due to finite-size effects, as we show in Appendix \ref{sec:finite_size}. 

If the theoretical partition function $F(x)$ used to compute $\la_n^\loc$ is the correct one, the 
$\Phi_K(\la_n^\loc)$ are uniformly distributed on the interval [0,1]. Thus, 
\bea \nonumber
\langle  \Phi_K \rangle &=& 0.5 \,, \\
\sqrt{\langle ( \Phi_K - \langle  \Phi_K \rangle)^2 \rangle} &=& 1/\sqrt{12} \approx 0.29\,,
\eea
where $\langle \cdot \rangle$ denotes an ensemble average and  we have used $\Phi_K \equiv \Phi_K(\la_n^\loc)$ for notational simplicity.
Assuming that the different realizations of $\Phi_K$ are statistically independent and using the central limit theorem, $\overline{\Phi_K} \equiv \frac{1}{m} \sum_{i=1}^m \Phi_{K,i}$ is then normally distributed around 0.5, with a standard deviation of $\sigma = \frac{1}{\sqrt{12 m}} \approx 0.0033$ for $m=7592$. 
However, note that this is not quite correct for $n=256$, due to finite-size effects, as we show in Appendix \ref{sec:finite_size}. We show with simulations that, for $n=256$, $\overline{\Phi_K} \equiv \frac{1}{m} \sum_{i=1}^m \Phi_{K,i}$ is distributed around $0.489$ rather than around $0.5$, i.e.\ it is biased low by about $3\,\sigma$. 
In Fig.~\ref{f:WMAPdecorPatch}, we also plot the histogram of $\Phi_K$ for all three bands. We do see the bias towards small $\Phi_K$ from the finite-size effects.
The values for the empirical mean and standard deviation of $\Phi_K$ are summarized in the top panel of 
Table \ref{t:la}. The results are consistent with theoretical expectations.


\subsubsection{Results: correlation method}\label{sec:corr}

Instead of de-correlating the data from the CMB map, we now directly compare the empirical partition function of
$\la_n^\loc$ for the original correlated data with our simulated partition functions $\Phi(n,\la)$ for correlated data. 
We show a map of $\la_n^\loc$ (nside=64) in Fig.~\ref{f:kolmap}. Except for the Galaxy, the structures that we see in this map simply reflect correlations in the CMB. This can be seen from the fact that these structures go away when we de-correlate the data (cf. left panel of Fig.~\ref{f:kolmap}), leaving us with white noise. The Galactic plane is visible in the de-correlated map as well, indicating that we are seeing radiation from the Galaxy.

In order to compare the consistency of the results with simulations, we go back to working with ring segments.
As compared to the simulations in Section \ref{sec:mock}, we modify our
analysis for data that do not lie on a great arc, but on a circle in a plane
parallel to the Galactic plane, characterized by an angle $\Theta_\ring$ in
Galactic coordinates. With this we can directly use the data
from the WMAP temperature maps in the HEALPix `ring' format. This has the
advantage 
that we do not need to rotate the map in order to obtain data on a great arc
that lies well above the Galactic plane (rotating the maps introduces
significant uncertainties in the data), while the simulation of the partition function is still easy to do.

From these rings, we choose segments of an opening angle
of $20^\circ$ and a given gap angle between the segments, in order to reduce
correlations between them (both angles are measured with respect to the centre
of the coordinate system). 
We choose the gap between the ring segments and between different rings to be $1^\circ$ each. In Appendix \ref{s:res_corr}, we verify with simulations that for such a gap we do not yet see any effects from correlations between the different ring segments.\footnote{We expect correlations between ring segments to manifest themselves in too steep curves for $\Phi(n,\la)$: 
Correlated ring segments `know about each other', which should make the $\la_n$ for the different ring segments more 
similar to each other, and thus the cumulative distribution for the $\la_n$ steeper. Note that since the patches are much better localized than the ring segments, they should be much less affected by this effect.}

In order to obtain as many ring segments as possible, we use all
rings between $\Theta_\ring = 20^\circ - 60^\circ$ with a distance of $1^\circ$.
We discard all pixels with point sources masked by the WMAP extended mask. If in a ring segment more than 
 10\% of the pixels are masked, we discard the whole ring segment, thus obtaining $m=712$ ring segments.
Due to the pixelisation scheme, the number of pixels contained in a given opening 
angle depends on the ring. For an opening angle of $10^\circ$, the number of pixels 
we use varies between $n=132$ and $n=150$ for $\Theta_\ring = 20^\circ - 60^\circ$. Furthermore, the masking of the pixels changes the number of pixels in a given ring segment.
 The variation in $n$ causes small differences between the 
simulated partition functions of the $\la_n$ for the `correlation method' introduced below. However, we have verified 
numerically that this effect is negligible compared to the variation we obtain due to the finite number of ring segments 
we can extract from the map. For the simulations, we use the average number of pixels in the ring segments, $n = 143$.

\begin{figure}
\centering
\includegraphics[width=7.5cm]{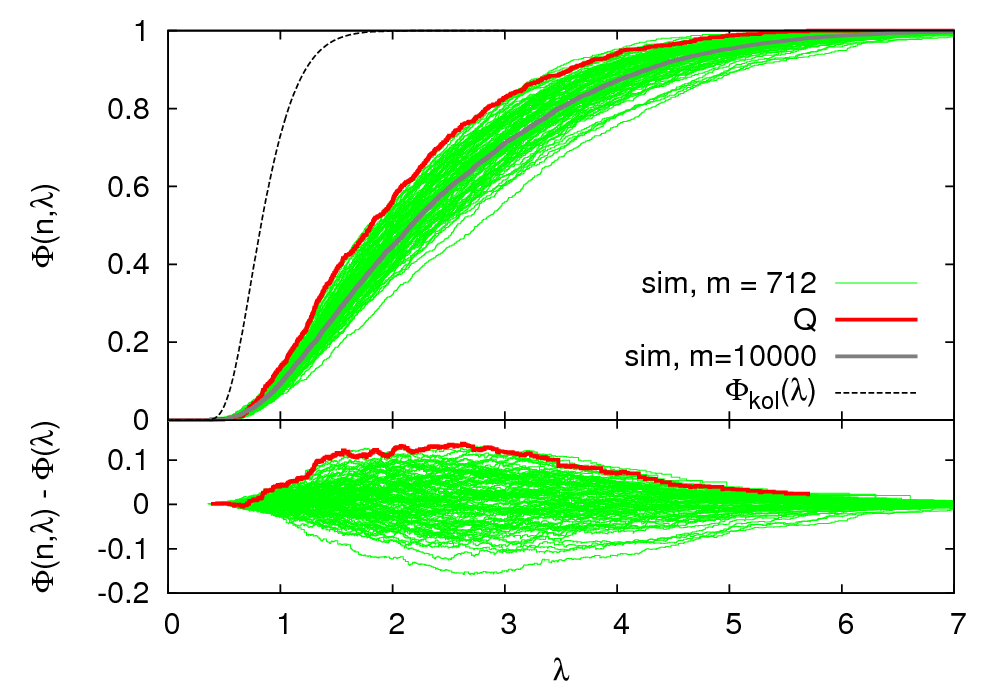}
\includegraphics[width=7.5cm]{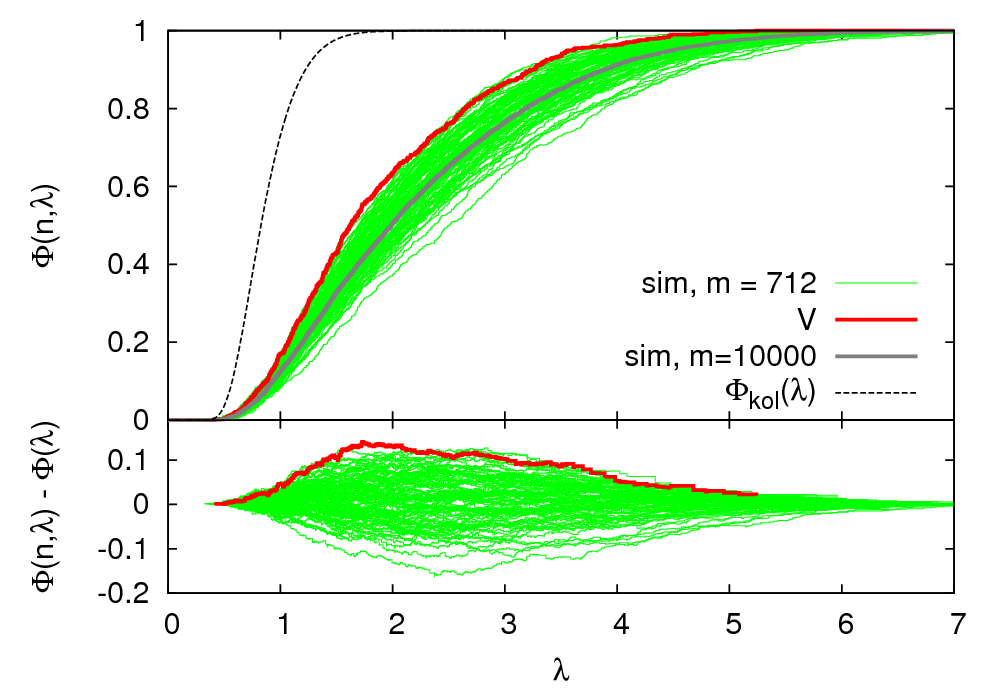}
\includegraphics[width=7.5cm]{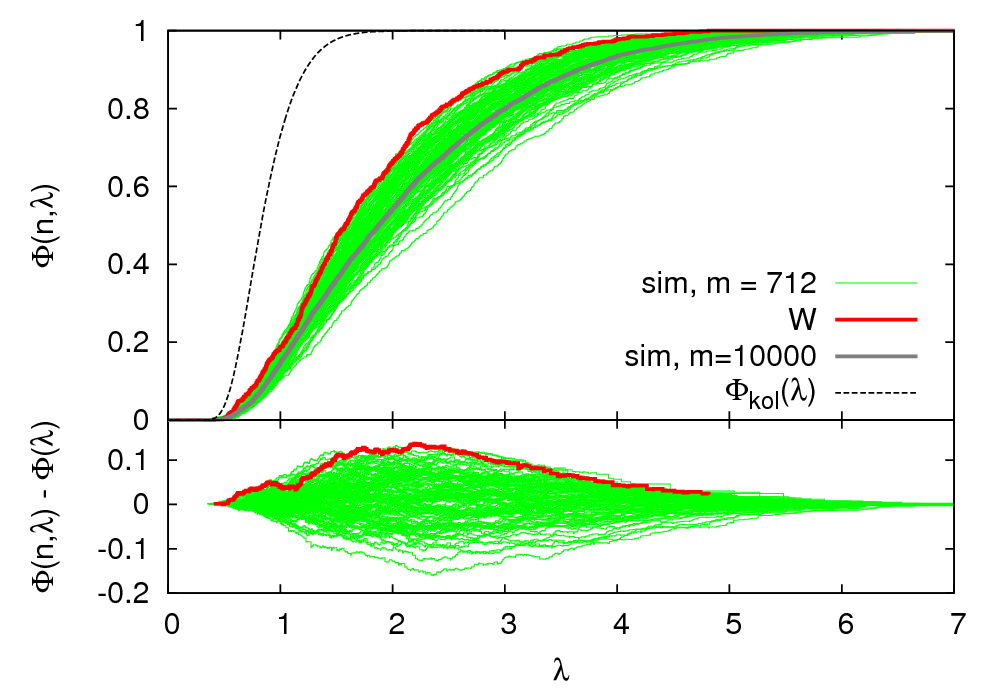}
\caption{Correlation method: empirical partition function of $\la_n^\loc$ for WMAP-7year-data for the Q-, V-, and W-band (upper left, upper right, and bottom panel, respectively). We compare these to simulated partition functions $\Phi(n, \la)$, obtained from m = 10000 samples, (grey curve) and 100 realizations of $n = 143$, $m = 712$ (green curves). 
The opening angle of a given ring segment is $20^\circ$, the gap between 
rings and ring segments are $1^\circ$ each. Pixels containing 
identified point sources are masked, and we take into account only homogeneous noise. The Kolmogorov-curve $\Phi_K(\la)$ is indicated by the black dashed line.}
\label{f:WMAPcor2}
\end{figure}

\begin{table}[]
\centering
\begin{tabular}{| c |  c  c  c  |}
\hline
band & Q & V & W \\
\hline
mean[$\Phi_K(\la_n^\loc$)] & 0.90 & 0.88 & 0.86 \\  
\hline
stddev[$\Phi_K(\la_n^\loc)$] & 0.19 & 0.22 & 0.25 \\   
\hline\end{tabular}
\caption{\label{t:laCorr} 
Results for the three WMAP bands for the correlation method for $m=712$ samples ($n \approx 143$): shown are the empirical mean and standard deviation of $\Phi_K(\la_n^\loc)$ over all ring segments. Note that now we obtain larger means for $\Phi_K(\la_n^\loc)$, simply because the $\la_n^\loc$ are larger due to the correlations between the pixels in a given ring segment. The standard deviation is smaller than 0.29, due to the flatness of the Kolmogorov-curve for large values of $\la_n$.
} 
\end{table}

We now consider only homogeneous noise. The effect of using the wrong noise variance in the 
correlation method actually introduces a bias, whereas in the de-correlation method it merely increases the scatter of 
the simulations. We have studied this effect by
excluding the regions where the noise is most inhomogeneous, i.e.\ the regions around the ecliptic poles, from the 
analysis.\footnote{Around the ecliptic poles, the WMAP scanning strategy affects the structure of the noise most 
strongly.} 
This left us with $m=357$ ring segments only, rather than $m=712$ as in the de-correlation method, thus 
worsening the statistical power. We
found that the bias is not significantly removed by this and therefore included also the 
pole region in the final analysis. 

The results for the different bands are shown in Fig.~\ref{f:WMAPcor2}. The curves obtained from the WMAP data are still marginallly consistent with the simulations, but they are now systematically too steep. This is most probably due to the fact that we only considered homogeneous noise in this analysis, while taking into account the correct noise covariance in the de-correlation method.\footnote{When considering only homogeneous noise in the de-correlation method, we also obtain curves which are less consistent with simulations than when taking into account the full noise covariance.}

In Table \ref{t:laCorr}, we show the empirical mean and standard deviation of $\Phi_K(\la_n^\loc)$, where $\Phi_K$ is 
the 
Kolmogorov distribution, and not the correct partition function for the correlated case. As expected, we obtain $\overline{ \Phi_K(\la_n^\loc)} > 0.5$, simply because the $\la_n^\loc$ are pushed towards higher values by the correlations within the ring segments. For the typical values of $\la_n \sim 0.9$ we obtain, the Kolmogorov-curve is already quite flat. Thus, $\Phi_K(\la_n^\loc)$ varies little over the ring segments, leading to a standard deviation of $\Phi_K$ which is smaller than $0.29$.


\subsection{WMAP - global analysis}\label{s:global}

In order to find out whether the CMB map as a whole is Gaussian distributed and described by the correlation function 
 used, we compute a global $\la_n$ of the CMB map: We randomly draw $n=2000$ pixels 
outside the mask from the map, de-correlate all of these pixels with their full covariance matrix (using inhomogeneous 
noise), and compute the global $\la_n$ from the resulting uncorrelated data. We compute $m=500$ of these global $
\la_n$.
The values for $n$ and $m$ where chosen to maximize the sensitivity of $\la_n$ to point-source contamination, while 
still being safe from residual correlations between the different sets of pixels, as we explain in Section \ref{s:ps}.

\begin{figure}
\centering
\includegraphics[width=7.5cm]{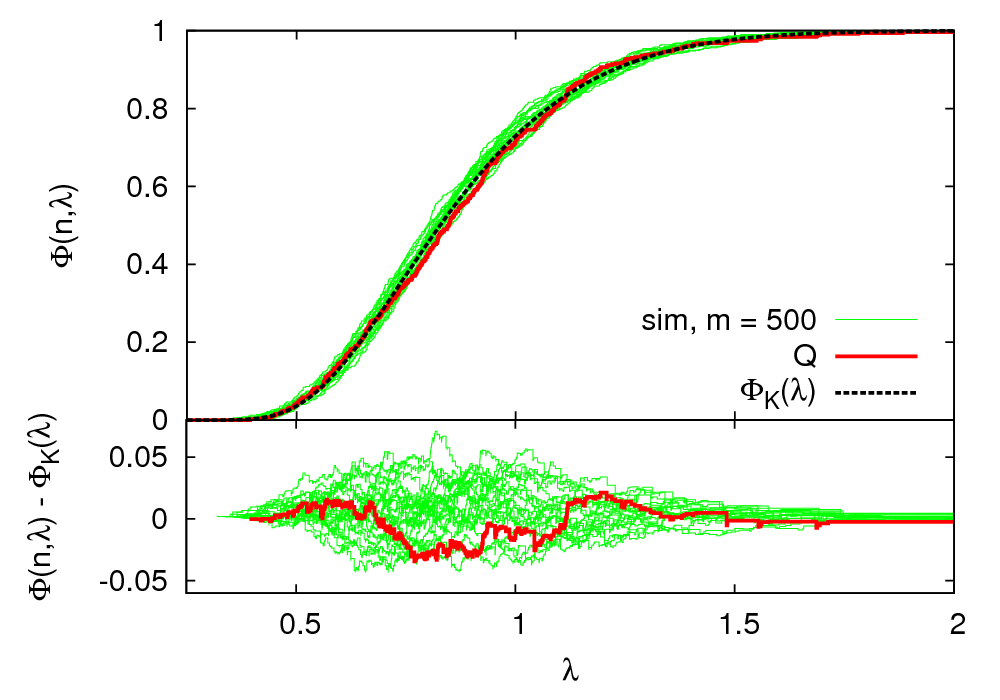}
\includegraphics[width=7.5cm]{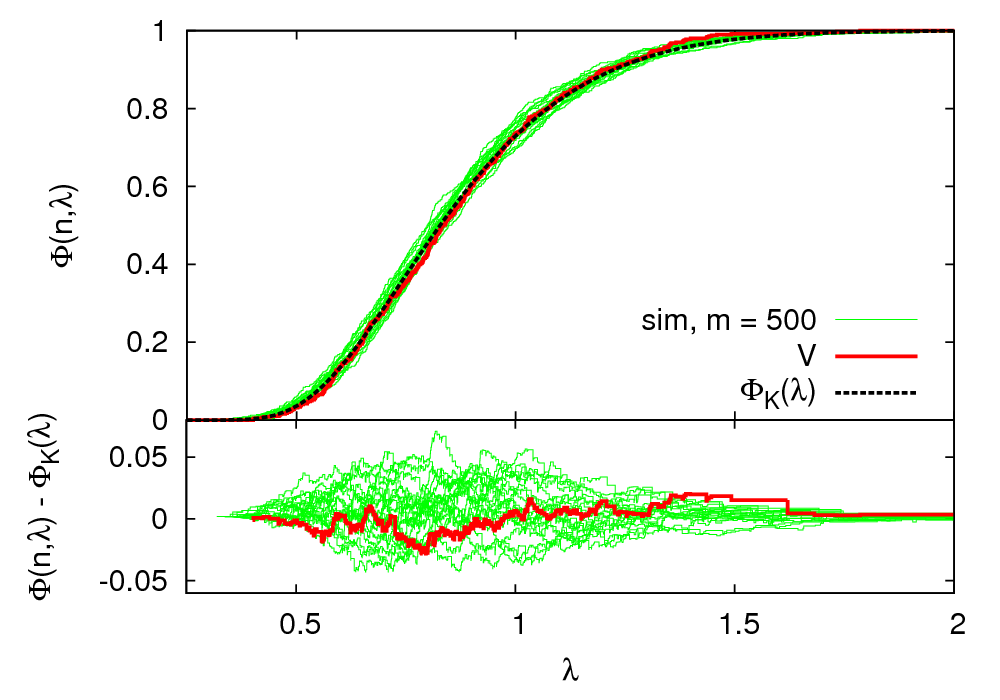}
\includegraphics[width=7.5cm]{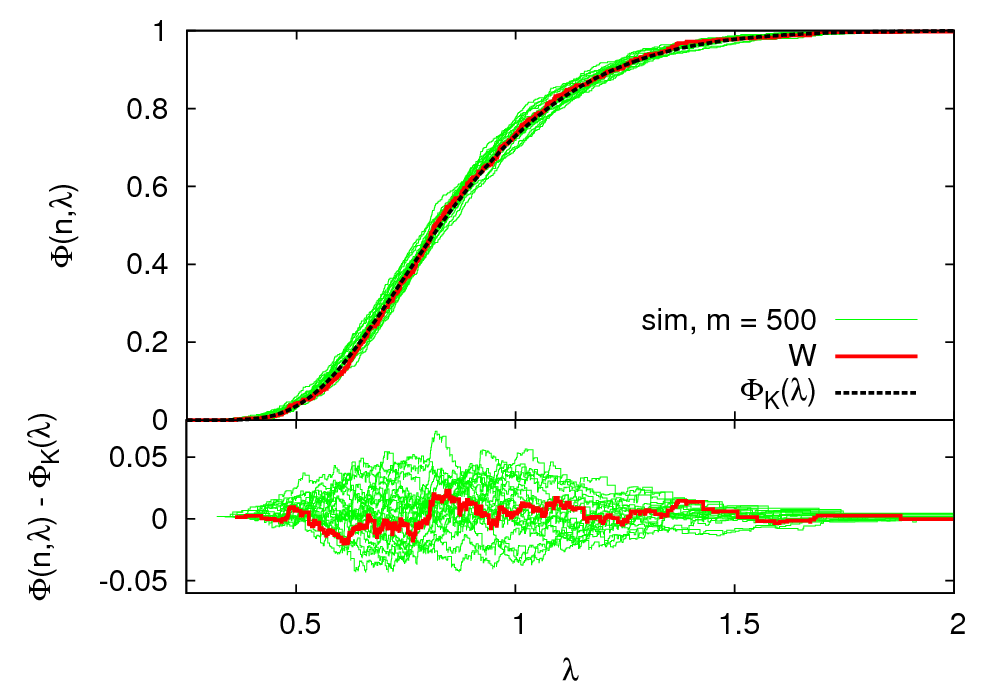}
\includegraphics[width=7.5cm]{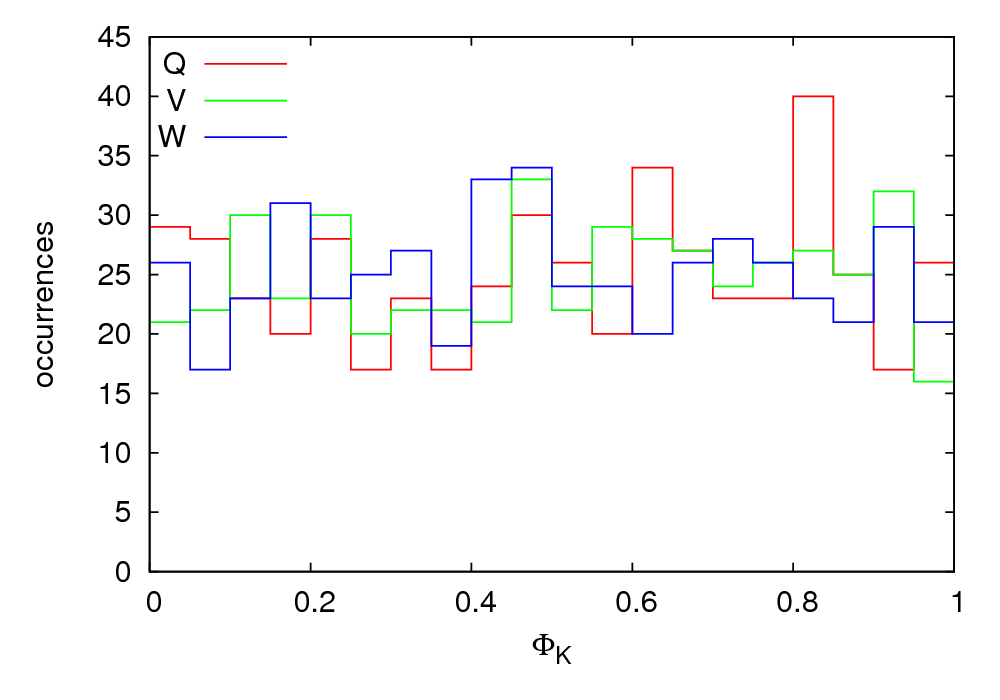}
\caption{Global analysis: empirical partition function of $\la_n^\loc$ for WMAP-7year-data and 20 simulations for $m=500$ sets of $n=2000$ randomly drawn pixels. \emph{Upper left:} Q-band, \emph{upper right:} V-band, \emph{bottom left:} W-band. \newline
\emph{Bottom right panel:} Histogram of $\Phi_K$ for the three bands. 
}
\label{f:distrLambda}
\end{figure}

In order to get a visual impression of the results, we show the empirical partition function of $\la_n$, together with 20 simulated partition functions in Fig.~\ref{f:distrLambda}. The WMAP-curves lie well within the range expected from simulations.

Just as for the local analysis in Section \ref{s:decorr}, the $\Phi_K(\la_n)$ have to be uniformly 
distributed on [0,1], if all our assumptions used to compute $\la_n$ are correct. 
We plot the histogram of $\Phi_K(\la_n)$ for $n=2000$ and $m=500$ for all bands in the bottom right panel of Fig.~\ref{f:distrLambda}.

In order to obtain a more quantitative result, we again compare the empirical mean and standard deviation to the theoretical expectations, c.f.\ Section \ref{s:decorr}.
Taking into account the very small bias from the effect of a finite $n$ (c.f.\ Appendix \ref{sec:finite_size}),
we expect $\overline{\Phi_K}$ to be normally distributed around $0.495$ with a standard deviation of $\sigma \approx 0.013$. The empirical standard deviation of $\Phi_K$ should be about $0.29$.
We summarize 
the empirical mean and standard deviation of $\Phi_K(\la_n)$ in Table~\ref{t:la}. The results are again fully consistent with theoretical expectations.

\section{Constraints on residual point sources}\label{s:ps}

\begin{figure}
\centering	
\includegraphics[width=7.5cm]{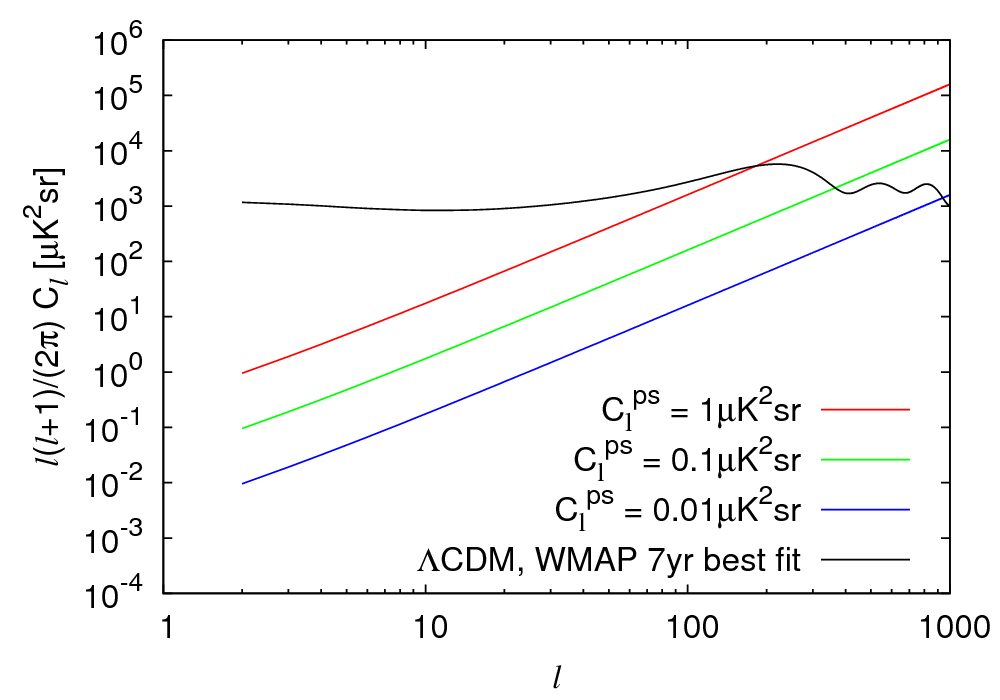}
\caption{CMB power spectrum and 3 different constant power spectra, describing the residual point source background. The WMAP-constraints on the power spectrum of residual point sources is of the order of $0.01\mu K^2 \sr$ or better.}
\label{f:clps}
\end{figure}

One can in principle use the Kolmogorov-Smirnov test to constrain the power coming from residual (non-resolved) point 
sources. We consider only radio point sources, which are spatially uncorrelated (white noise) and thus have a flat 
power spectrum.
Toffolatti et al.~\cite{Toffolatti1998} show that the residual radio point sources are well described by a Poisson 
distribution. In this work, we approximate this Poisson distribution by a Gaussian, which is a good approximation if the 
Poisson distribution has a large mean. We can then simply 
add a constant contribution $C_\ell^\ps = {\rm const}$ to the CMB power spectrum and re-run the Kolmogorov-Smirnov test. We 
plot different point source power spectra together with the CMB power spectrum for the WMAP seven year best fit 
cosmological parameters in Fig.~\ref{f:clps}. 

Using the spectral behaviour of the detected point sources, the frequency dependence of $C_\ell^\ps$ can be modeled 
as a power law \citep{Nolta2009}
\be
C_\ell^\ps(\nu) = A_\ps \, r(\nu)^2 \left( \frac{\nu}{\nu_Q}  \right)^{2\alpha-4} \,,
\label{e:constrWMAP}
\ee
where $\alpha = -0.09$, $\nu_Q = 40.7$~GHz is the frequency of the Q-band, and $r(\nu)$ converts antenna to thermodynamic temperature:
\be
r(\nu) \equiv \frac{(e^x-1)^2}{x^2e^x} \,,   \,\,\,\,\,  x \equiv \frac{h\nu}{k T_\cmb} \,.
\ee
The WMAP seven year estimate for the amplitude is $A_\ps = (9.0 \pm 0.7)\times 10^{-3} \mu K^2 \sr$~\citep
{Larson2011}. The constraints on $C_\ell^\ps$ on the different bands obtained from Eq.\ (\ref{e:constrWMAP}) given 
this result for $A_\ps$ are listed in Table \ref{t:ps}.

In order to obtain constraints on the residual point sources from the Kolmogorov-Smirnov test, we add a constant point-source 
power spectrum to the CMB power spectrum and compute again the $m=500$ values for the global $\la_n$ with 
$n=2000$, as in Section \ref{s:decorr}. 

As we have already stated above, the values for $n$ and $m$ where chosen to maximize the sensitivity of $\la_n$ to point-source contamination, while still being safe from residual correlations 
between the different sets of pixels:
We found that the sensitivity of $\la_n$ to a contamination of the CMB power spectrum by point sources is much higher 
for large $n$ than for small $n$. This can be understood as follows: The effect of adding point-source power to the 
CMB-power spectrum is to change the partition function $F(x)$ of our de-correlated pixels. This 
increases the value of $\sup_x|F_n(x)-F(x)|$, c.f. equation (\ref{def_la}). Now, since for large $n$ the original 
supremum (without point-source power added) is smaller than for small $n$, it feels the change of $F(x)$ stronger (i.e. 
its relative change is larger than for small $n$).
We thus chose to set $n=2000$, for which we can still diagonalize the covariance matrix in a reasonable amount of time. In addition, for $n=2000$ pixels, the effects from finite $n$ are already very small, c.f.\ Appendix \ref{sec:finite_size}.

The number of pixels in our map is $N_\pix = 12 \, \nside^2 \sim 3\times 10^6$ for $\nside = 512$. With roughly 30\% of 
the sky masked, we are left with $\sim 2\times10^6$ pixels outside the mask. Therefore, every set of $n=2000$ pixels 
we draw represents about 0.1\% of the pixels outside the mask. If we draw $m=500$ such pixel-sets, we have in total 
used about 50\% of the pixels outside the mask (neglecting the fact that some of those might have been drawn several 
times). We show in Appendix \ref{s:res_corr} that with this we are safe from effects from residual correlations between the different  pixel sets.

\begin{figure}
\centering	
\includegraphics[width=7.5cm]{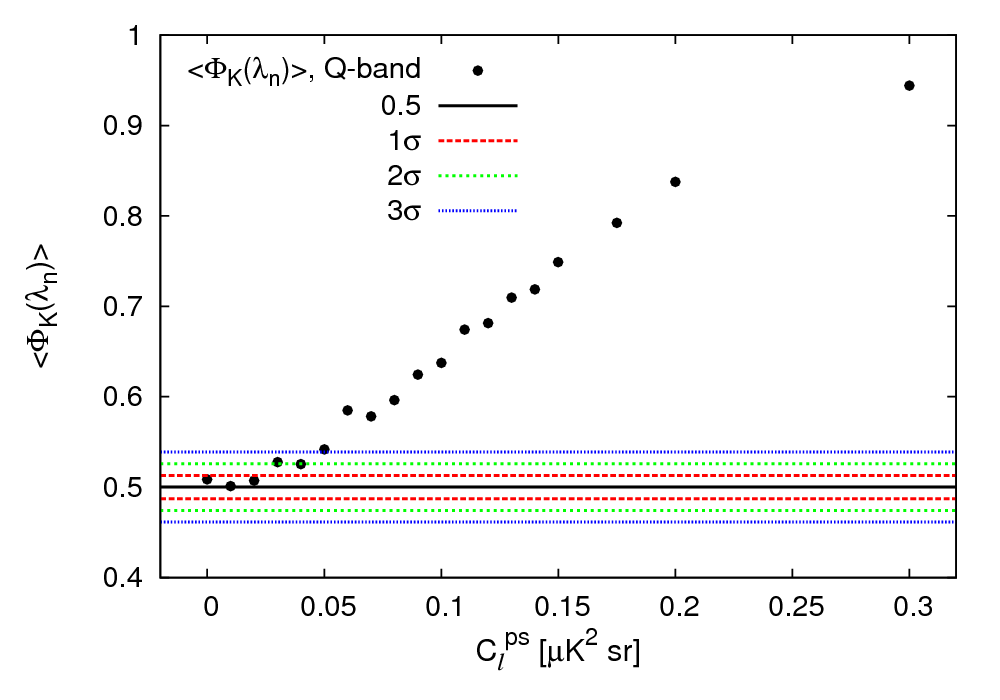}
\includegraphics[width=7.5cm]{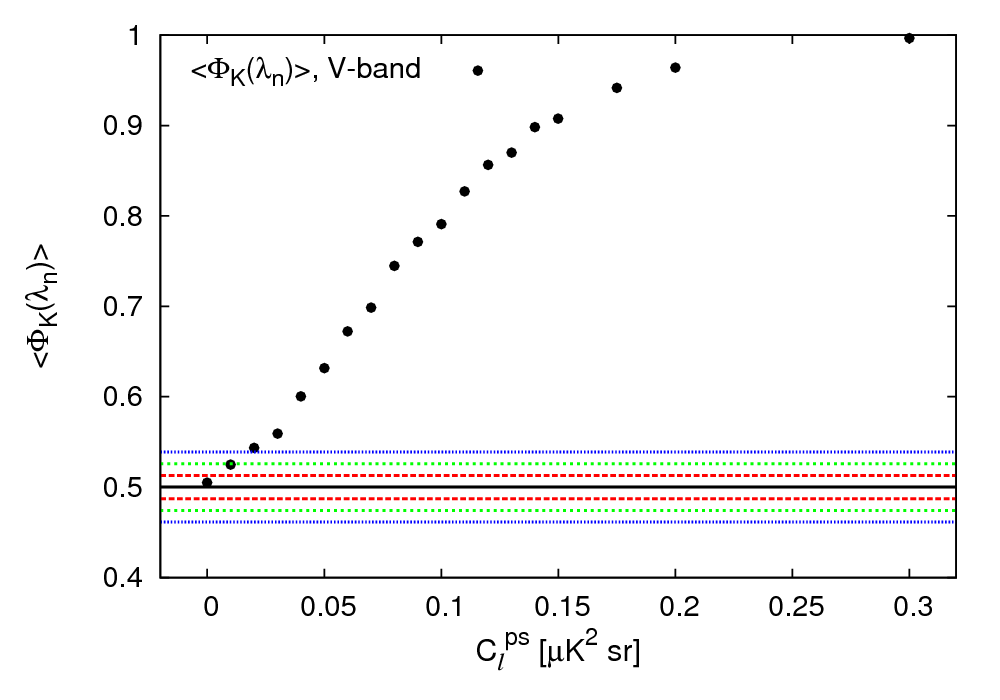}
\includegraphics[width=7.5cm]{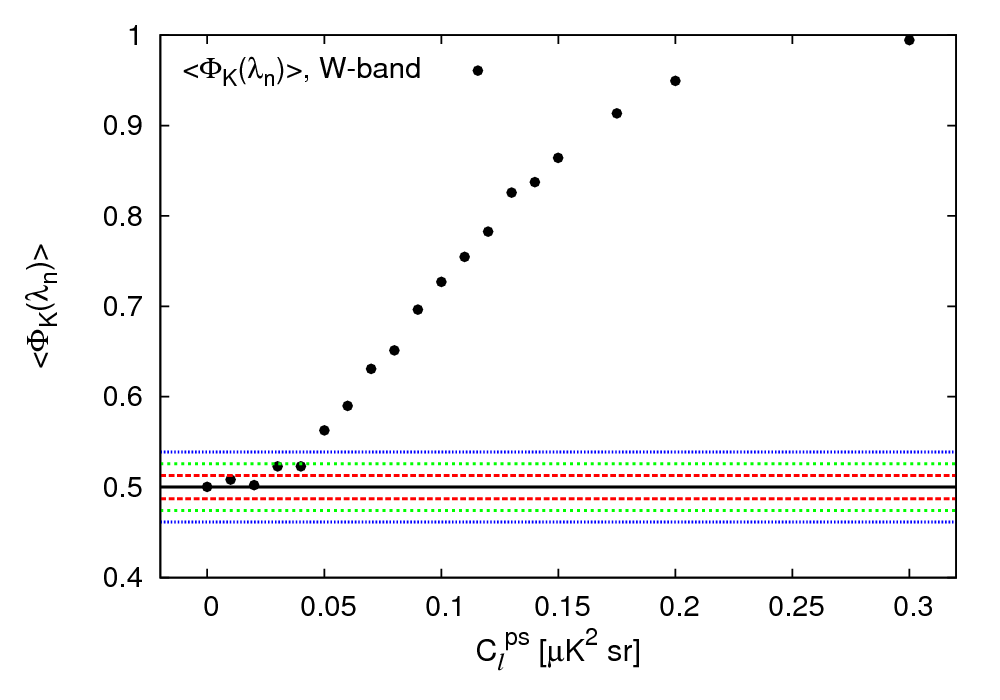}
\caption{Empirical mean of $\Phi_K(\la_n)$ versus the Poisson power $C_\ell^\ps$ for the Q-band (\emph{upper left}), V-band 
(\emph{upper right}), and W-band (\emph{bottom}), for n=2000 and m=500. The confidence intervals around the expected value of 0.5 are also shown. }
\label{f:constraints}
\end{figure}

As in Section \ref{s:global}, we again use the fact that the empirical mean $\overline{\Phi_K}$ has to be normally distributed around 0.5 with a standard deviation of $\sigma \approx 0.013$ if we used the correct partition function $F$ to compute $\la_n$.\footnote{Note that we neglect the small bias from finite $n$ here. This results in slightly more conservative constraints than with the bias taken into account, since $\overline{\Phi_K}$ is biased towards smaller values.} If we add too much point-source power, this hypothesis is no longer true, and $\overline{\Phi_K}$ can become very different from 0.5. 
 We thus add point-source power $C_\ell^\ps$
of different strengths to the CMB power spectrum, and plot the resulting $\overline{\Phi_K}$ versus 
$C_\ell^\ps$ in Fig.~\ref{f:constraints}. The $1\sigma$, $2\sigma$, and 
$3\sigma$ regions around the expected value of 0.5 are also shown. 
We obtain an upper bound on the point source power by rejecting the hypothesis if our measured value of $\overline{\Phi_K}$ lies outside the $3\sigma$ region. 
The bounds we obtain for every band are listed in Table~\ref{t:ps}.  The Q-band is affected 
 least by the point sources, since it has the lowest resolution, and thus the beam window 
function cuts the higher multipoles, when the point sources are more important as compared 
to the CMB.

\begin{table*}[]
\centering
\begin{tabular}{| c |  c | c | c  |}
\hline
band & Q & V & W \\
\hline
WMAP 7 yr & $(9.8 \pm 0.8)10^{-3}$ &   $(2.0 \pm 0.2)10^{-3}$  &  $(0.43 \pm 0.03)10^{-3}$ \\
\hline
Kol. (de-corr) & $<0.05$ & $<0.02$ & $<0.05$ \\
\hline
\end{tabular}
\caption{\label{t:ps} Constraints on $C_\ell^\ps$ in $\mu K^2 \sr$ from the WMAP team and from our Kolmogorov analysis.} 
\end{table*}

Our results obtained in this way are about $0.5-1$ order of magnitude worse than the WMAP
limits. However, note that the WMAP team has combined different bands and used an assumption about the scaling of $C_\ell^\ps$ with frequency to obtain their constraints. We do not use any assumption on this and derive independent constraints for each band.
We could most probably improve the constraints by using a higher $n$ for computing 
$\la_n$, thus enhancing the sensitivity of $\la_n$ to residual point-source power 
as explained above. However, as already stated above, with 
$n=2000$, we are already close to the limit of what we can do with the matrix 
diagonalization routine we are presently using. 
Therefore, working with a larger $n$ would involve using more sophisticated routines and is beyond the scope of this work.
Note also that our limits are quite conservative, given that they were obtained using only 50\% of the pixels
outside the mask, i.e. about 35\% of the sky. 
We could also improve our constraints by using more than 500 samples, but we would run the risk of working with correlated samples (if we were to draw the additional samples outside our current mask), or of being contaminated by Galactic foregrounds (if using a smaller mask). We prefer to be safe from these effects and thus stick to the more conservative estimate. 
Finally, note that we have chosen to reject the hypothesis if the data lie outside the 
$3\sigma$ region around the expected value, which is again quite conservative.


\section{Conclusion and outlook}\label{s:con} 

In this paper we have shown that the Kolmogorov stochasticity parameter of the
correlated data $\De T(\bn)$ does not obey a Kolmogorov distribution and we 
have explained why this is to be expected.
We have also presented a method to de-correlate the data so that the Kolmogorov-Smirnov 
test on the de-correlated data should give an indication whether or not the
CMB anisotropies are Gaussian. Alternatively, we can apply the Kolmogorov-Smirnov test directly to the correlated data, but with a modified partition function, which takes into account the correlations and which is obtained from simulations. This latter method is to be favoured over de-correlating the data in the case of very strongly correlated data.

We have applied the Kolmogorov-Smirnov test on WMAP seven year data using both of these methods, and we have shown that the WMAP maps are fully consistent with a Gaussian dataset with the correlation function given by the best-fit cosmological parameters.

We have then used the Kolmogorov-Smirnov test to obtain conservative constraints on the constant power spectrum of residual radio point sources. With the numerical effort involved and the present resolution of the maps, our results
are not competitive with the standard analysis but we are convinced that this method has
considerable potential.

Given that the spherical harmonics coefficients $a_{\ell m}$ of an isotropic sky map are uncorrelated, they are probably a
better choice of variables to constrain residual point source power in the map. 
However, this is beyond the scope of the present work, and presumably mostly useful 
for higher resolution maps as we expect to obtain from the Planck satellite currently 
taking data. These maps may make this method very competitive
and certainly an interesting alternative to the standard analysis. 

The Kolmogorov-Smirnov test can also be used to detect Galactic foregrounds, such as synchrotron radiation or dust. This should of course be done locally in pixel space.

One may also use the fact that the $a_{\ell m}$'s should be independent Gaussian variables with variance
$C_\ell$ to determine cosmological parameters. For a given model determined by a list $A$ of cosmological parameters, e.g., $A= (n_s, \De_R, \Om_ch^2,\Om_bh^2,\Om_\La,\tau)$ we can test whether the ratio
of observed $a_{\ell m}$'s and the theoretical $C_\ell$'s are independent with distribution $N(0,1)$.
More precisely, the quantities
$$ 
\frac{{\rm Re}[a_{\ell m}]}{\sqrt{C_\ell/2}} \qquad \mbox{ and } \qquad \frac{{\rm Im}[a_{\ell m}]}{\sqrt{C_\ell/2}}
$$ should be distributed with $N(0,1)$.
One may define a Markov Chain Monte Carlo (MCMC) algorithm which minimizes the distance of the distribution 
of the resulting Kolmogorov stochasticity parameter with the Kolmogorov distribution. It is easy to obtain 
the best fit in this way, but to determine the errors will probably require some thought. A simple first guess 
is to use least square deviation of the functions $\Phi(\la,A)$ and $\Phi_K(\la)$. This corresponds to assuming Gaussian errors and may not be so bad if one has sufficiently many points $\la$, due to the central limit theorem. The width of the distribution, $\De\Phi(\la)$ can be determined by simulations using the experimental errors.
Even if this method might turn out to be slower than the MCMC algorithms presently in use, it provides an independent alternative which always has its merits.

Finally, one may use the normalized $a_{\ell m}$'s to test for non-Gaussianity. If even for the best fit 
cosmological parameters, the distribution of the Kolmogorov stochasticity parameter deviates from 
the Kolmogorov distribution, this may indicate that the  $a_{\ell m}$'s are not Gaussian distributed.
These possibilities will be explored in detail in a future project.

\section*{Acknowledgment}
We thank Martin Gander,
Martin Reinecke, Thomas Riller for guidance and help with numerical aspects of the 
project.  We are grateful  for discussions with Jean-Pierre Eckmann and Martin Kunz and thank the anonymous referee for useful suggestions. We acknowledge the use of HEALPix~\cite{healpix}, 
cmbeasy~\cite{cmbeasy}, and the WMAP data \footnote{http://lambda.gsfc.nasa.gov/product/map/current/m\_products.cfm}. This work is supported financially by the Fonds National Suisse.

\begin{appendix}

\section{De-correlation method applied to ring segments}\label{app:decorr}

\begin{figure}
\centering
\includegraphics[width=7.5cm]{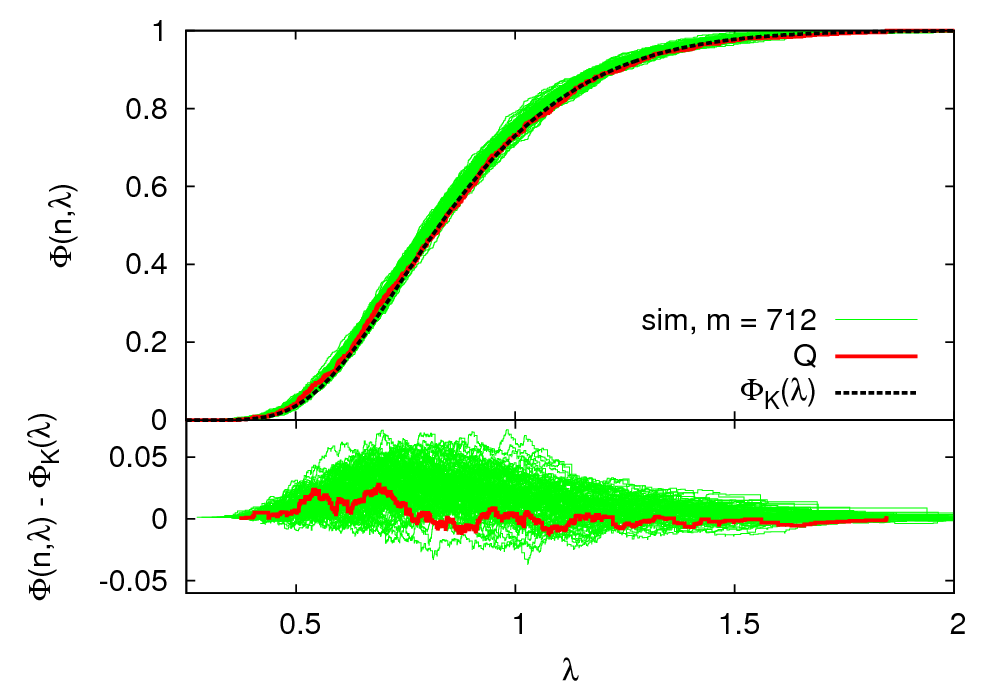}
\includegraphics[width=7.5cm]{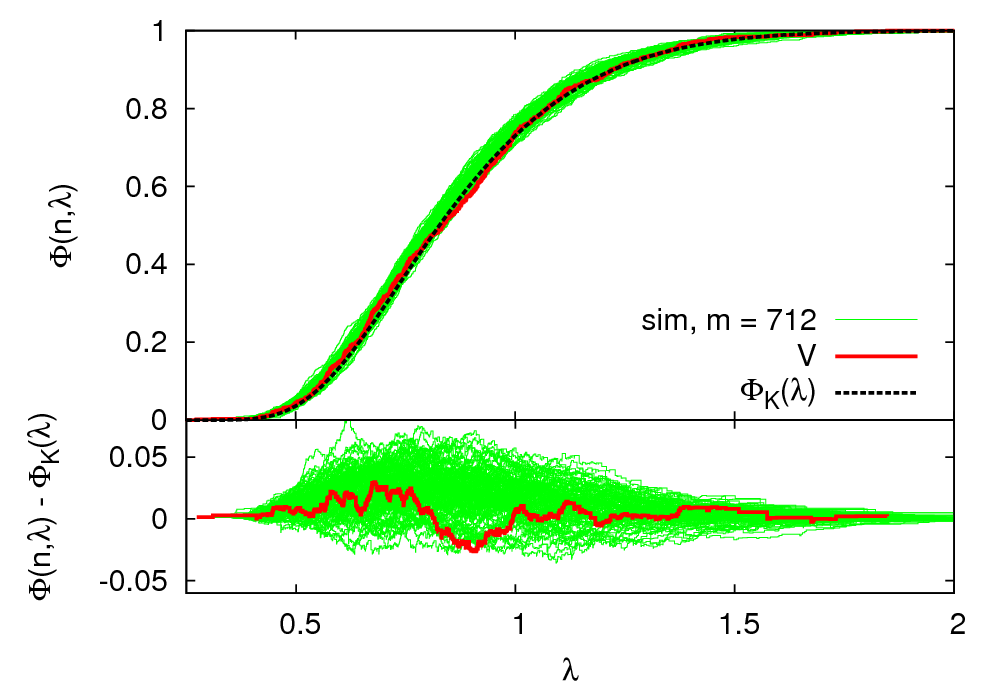}
\includegraphics[width=7.5cm]{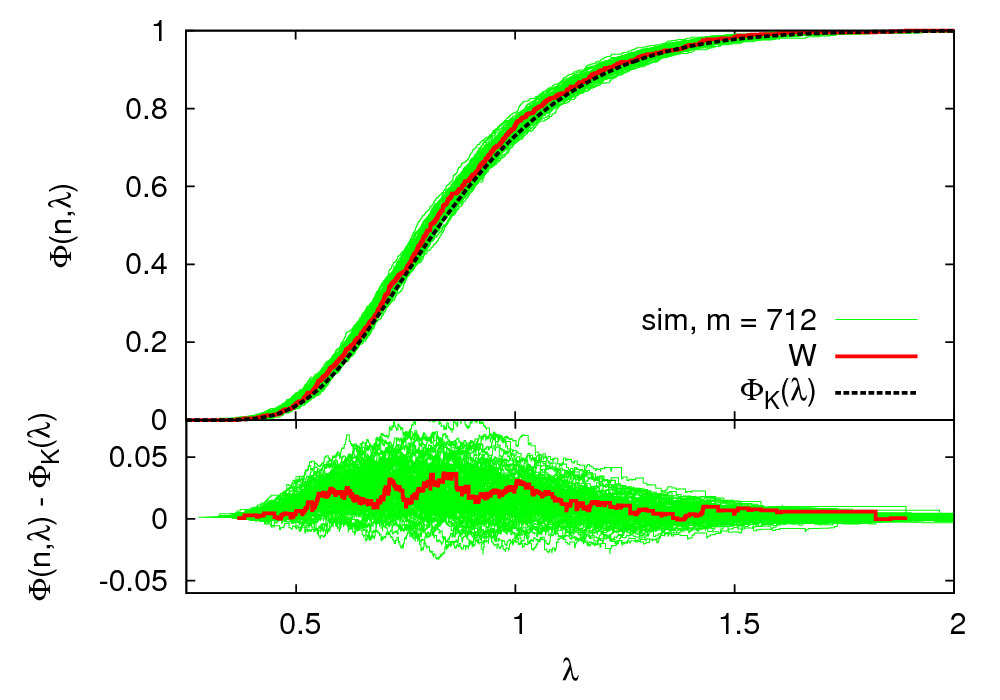}
\includegraphics[width=7.5cm]{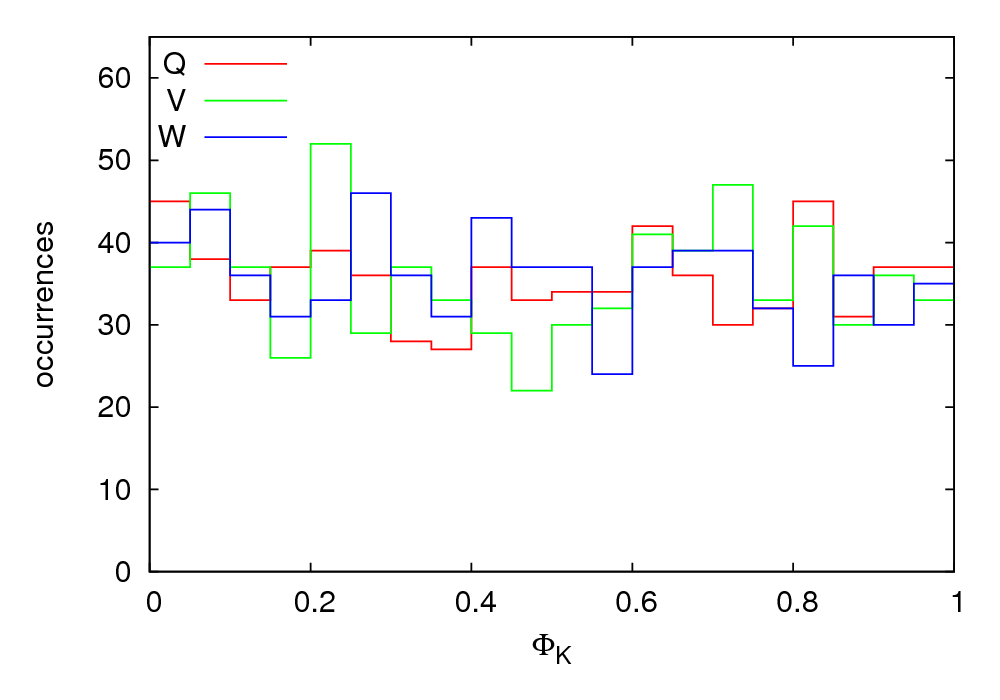}
\caption{Decorrelation method: empirical partition function of $\la_n^\loc$ for WMAP-7year-data and 100 simulations 
for the Q-band (\emph{upper left}), the V-band (\emph{upper right}), and the W-band (\emph{bottom left}).
The opening angle of a given ring segment is $20^\circ$, the gap between 
rings and ring segments are $1^\circ$ each. Pixels containing 
identified point sources are masked. We have $m=712$ ring segments and on average $n=143$ pixels per ring segment. The simulations are shown in green. \newline
\emph{Bottom right:} histogram of $\Phi_K$ for all bands. We see a small bias towards small values of $\Phi_K$, reflecting the effects of small $n$.}
\label{f:WMAPdecor}
\end{figure}

In order to easily compare the results of the de-correlation method to the ones obtained with the correlation method, we redo the de-correlation analysis on the ring segments used in Section \ref{sec:corr} and show the results in Fig.~\ref{f:WMAPdecor} and in Table~\ref{t:laA}. Taking into account the finite-size effect, we expect the empirical mean of $\Phi_K$ to be Gaussian distributed around $\langle  \Phi_K \rangle = 0.484$ with a standard deviation of $\sigma \approx 0.011$ (cf. Appendix \ref{sec:finite_size}). The results are fully consistent with theoretical expectations and, in contrast to the results from the correlation method, not biased. As already mentioned above, we think that the bias in the correlation method comes from the fact that we do not account for the correct (inhomogeneous) noise, but use homogeneous noise in the simulations.

\begin{table}
\centering
\begin{tabular}{|c| c |  c  c  c  |}
\hline
method &  & Q & V & W \\
\hline
\hline
de-correlation (ring segments) & mean[$\Phi_K(\la_n^\loc$)] & 0.496 & 0.497 & 0.483 \\
\hline
 & stddev[$\Phi_K(\la_n^\loc)$] & 0.295 & 0.295 & 0.289 \\
\hline
\end{tabular}
\caption{\label{t:laA} Empirical mean and standard deviation of $\Phi_K(\la_n)$ of the de-correlation method applied to the ring segments with $n\approx 143$, $m=712$.
The empirical mean should be Gaussian distributed around $\langle  \Phi_K \rangle = 0.484$ with a standard deviation of $\sigma \approx 0.011$.
The empirical standard deviation should be about $\sqrt{\langle ( \Phi_K - \langle  \Phi_K \rangle)^2 \rangle} = 1/\sqrt{12} \approx 0.29$. The analysis is fully consistent with theoretical expectations.} 
\end{table}

\section{Effect of the empirical determination of the data variance for a $\La$CDM model}\label{sec:lilliefors}

The power spectrum used to de-correlate the data and to perform the simulations of the
 correlated data, has been  determined as the best-fit power spectrum for the 
 $\La$CDM cosmological model from the CMB maps themselves \cite{Larson2011}. 
 Therefore, we expect in principle that the $\la_n$ tend to be smaller as compared to 
 the case when we work with the correct theoretical power spectrum, describing the Gaussian 
 distribution function from which the data have been 
drawn. This should shift the partition function $\Phi(\la_n)$ to the left.
This fact is known under the name Lilliefors effect and a test taking this into account is 
the so called Lilliefors test~\cite{liliefors}.

In this Appendix, we investigate this effect and show that it is negligible in our study. We start by simulating a Gaussian CMB map from the WMAP best fit power spectrum, described by 
the parameters in Table~\ref{t:parms}. 

From this map, we extract the best-fit power spectrum for a $\La$CDM model, using a simple 
$\chi^2$ likelihood in $C_\ell$ and a simplex algorithm to find its maximum. The theoretical 
power spectra have been computed with cmbeasy~\cite{cmbeasy}.

\begin{figure}
\centering
\includegraphics[width=7.5cm]{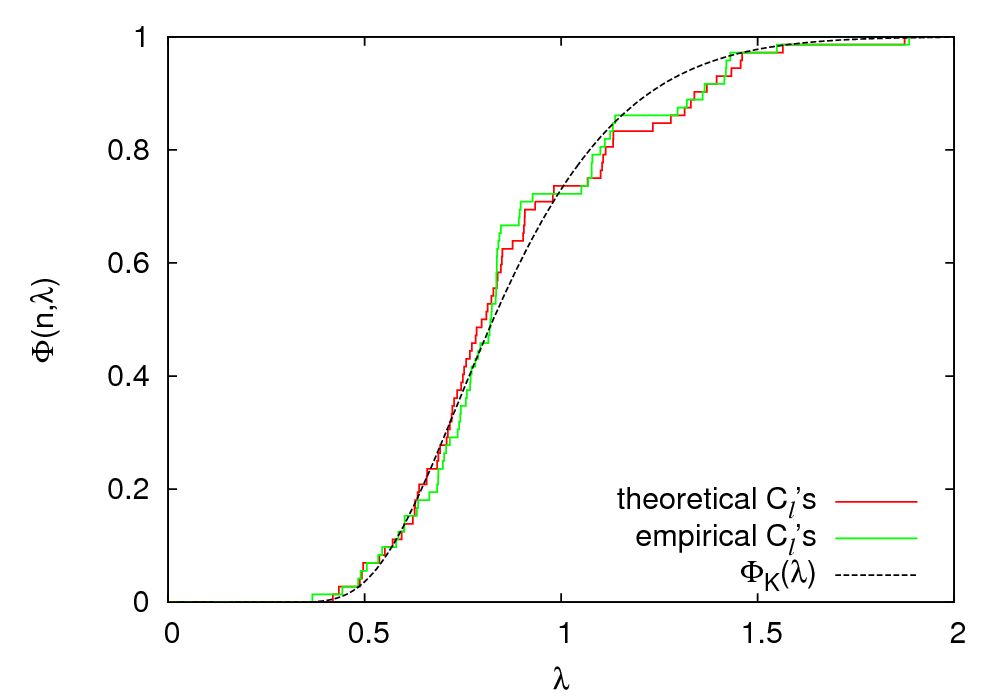}
\caption{Empirical partition function of $\la_n$ of simulated CMB data, de-correlated with the theoretical (i.e.\ correct) 
$C_\ell $'s (red curve) and the $C_\ell$'s from the best-fit $\Lambda$CDM model to the simulated data (green curve). }
\label{f:lil}
\end{figure}

We then de-correlate the simulated data once with the correct power spectrum and once with 
the best-fit power spectrum to the simulated data, and plot the empirical partition function of 
$\la_n$ for both cases. The results are shown in Fig.~\ref{f:lil}. There is only a very small
 difference between the two cases, so that we can safely ignore this effect in our analysis.

We assume that this is so because the derived distribution contains only six parameters (the cosmological parameters 
of the minimal $\Lambda$CDM model), which puts a strong prior on the correlation function, and because it has been derived using many more pixels than the ones used in the 
Kolmogorov-Smirnov test.

\section{Numerical issues for very strongly correlated pixels}\label{sec:pixelisation}

Very strong correlations between pixels can lead to numerical instabilities in the de-correlation method, as we show in the following. This can be of relevance if the noise in a CMB map is so low that it cannot make up for the missing power on small scales, which is cut by the beam of the experiment. We study these effects with simulated CMB data without noise. We first look at a Healpix-map of $\nside=512$. The latter has very small pixels ($0.013 \deg^2$) as compared to the WMAP beam ($0.048 \deg^2$ for the W-band and $0.26 \deg^2$ for the Q-band). Thus, neighbouring pixels are presumably strongly correlated, resulting in a covariance matrix with eigenvalues close to zero. We therefore expect to run into numerical problems as discussed in Section \ref{ss:dec}. Indeed, when we take into account the WMAP beam, we even obtain very small negative eigenvalues, which we think are due to numerical problems in the diagonalization routine. In this case, we cannot take the square-root of the eigenvalue 
and eq.~(\ref{e:decor}) becomes ill-defined. 

\begin{figure}
\centering
\includegraphics[width=7.5cm]{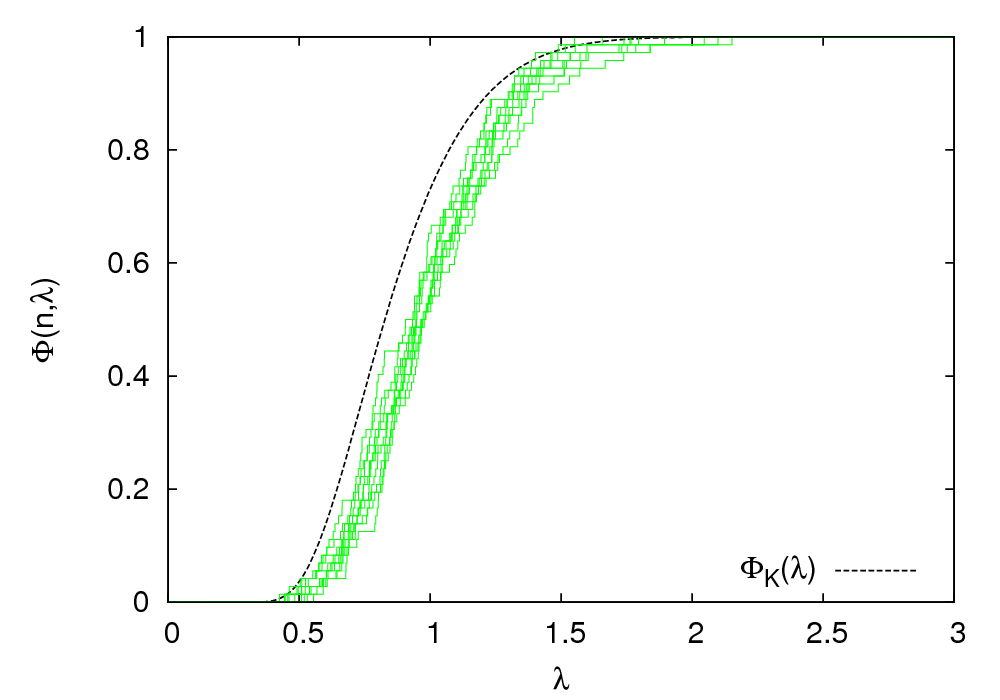}
\includegraphics[width=7.5cm]{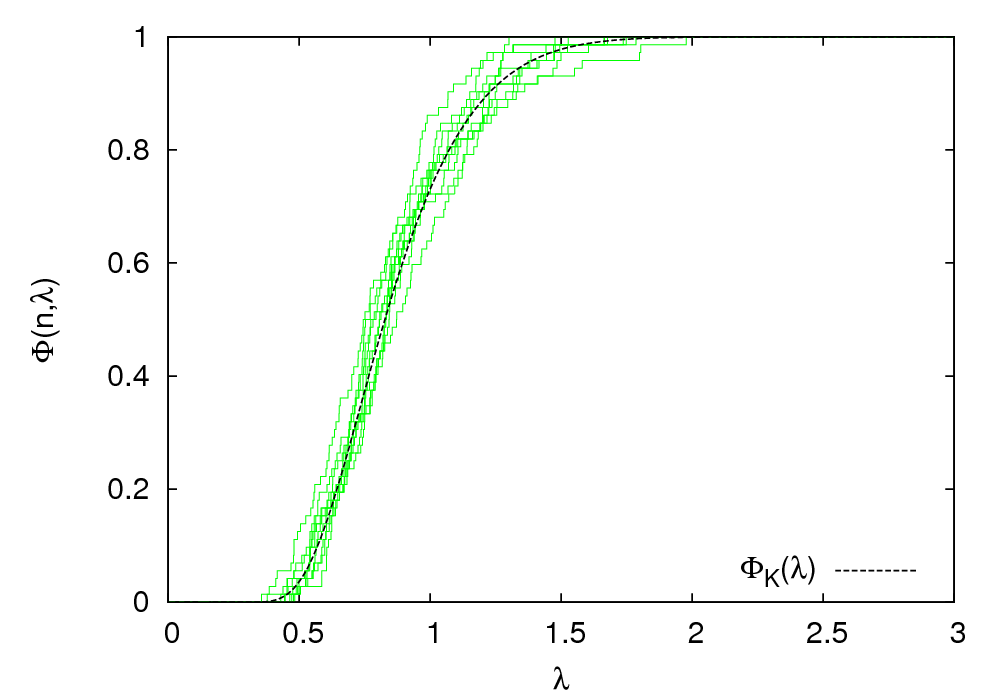}
\caption{Numerical issues: Empirical partition function of $\la_n$ using (very strongly correlated) neighbouring pixels (\emph{left}), and using 
every third pixel only (\emph{right}). When we use neighbouring pixels, the curves are biased to higher values of $
\la_n$.}
\label{f:pixelisation}
\end{figure}

Even if we do not take into account the beam, but we use all the power up to  $\ell_\Max = 1400$ for both the correlation function and the simulated maps, we do not get a meaningful result. The covariance matrix still has eigenvalues close to zero, which can lead to numerical instabilities in eq.~(\ref{e:decor}). We show the effect of this in Fig.~\ref{f:pixelisation}:
Using neighbouring (i.e.\ strongly correlated) pixels heavily biases the curves, whereas using only every third pixel yields unbiased curves. Not being aware of this effect can lead to wrong conclusions when interpreting the results of the Kolmogorov-Smirnov test. However, we have verified that this effect is no longer present when we add the detector noise of the WMAP experiment to the simulations. 

\section{Finite-size effects}\label{sec:finite_size}

\begin{figure}
\centering
\includegraphics[width=7.5cm]{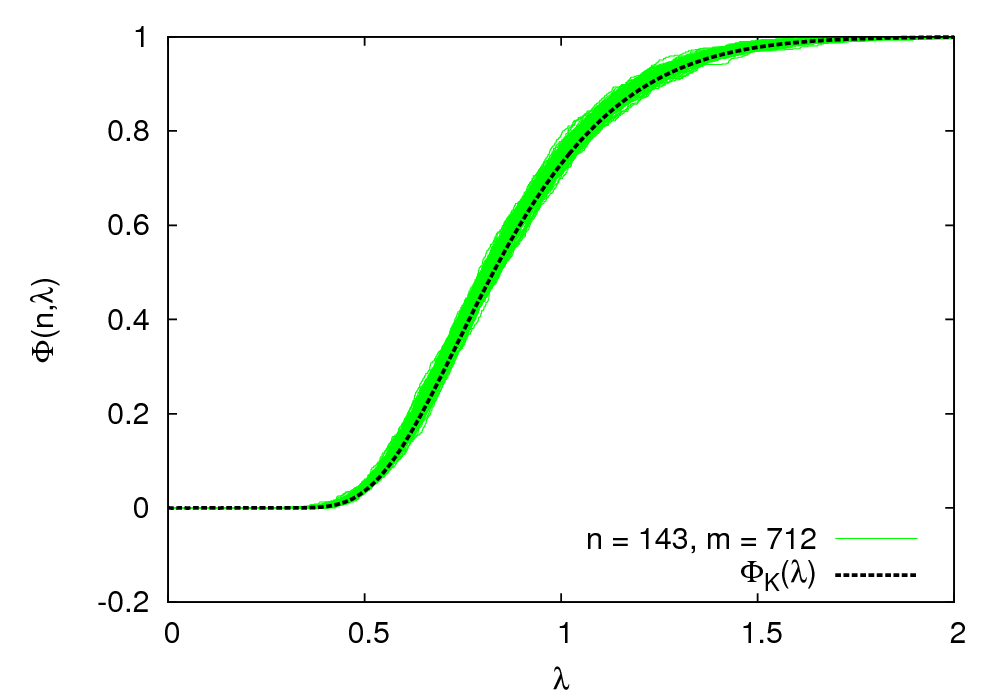}
\includegraphics[width=7.5cm]{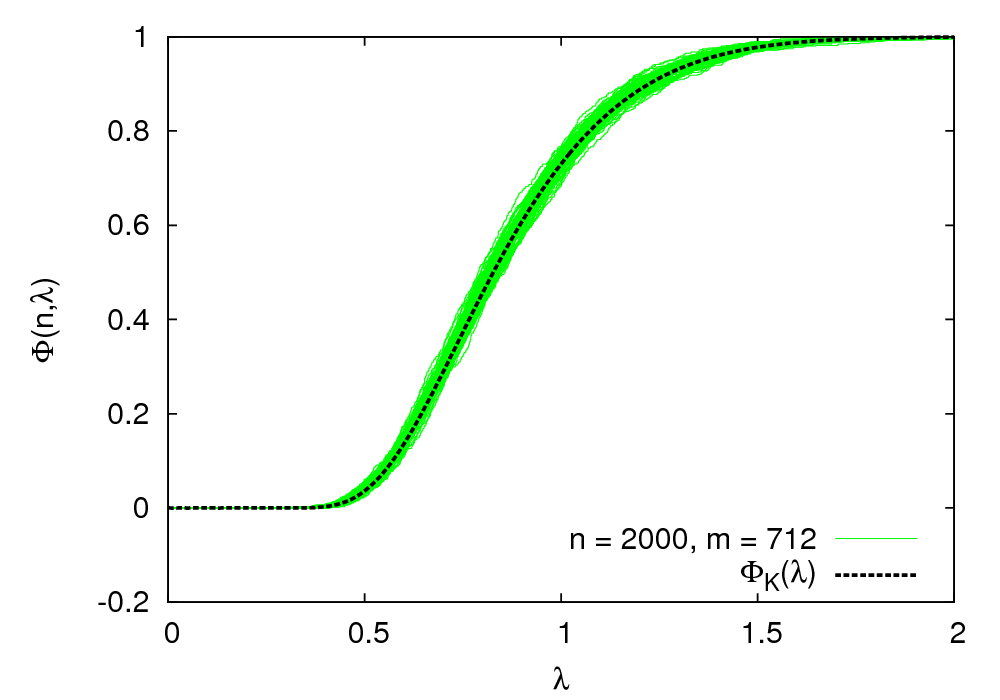}
\caption{Investigation of effects from small $n$: \emph{Left panel:}
100 realizations of the empirical partition function of $\la_n$ for $m=712$ independent ring segments of $n=143$ pixels each. \newline
\emph{Right panel:} Same for for $m=712$ independent ring segments of $n=2000$ pixels each. The curves for $n=143$ lie mostly to the left of the Kolmogorov-function, whereas the ones for $n=2000$ scatter more symmetrically around the latter.}
\label{f:bias}
\end{figure}

\begin{figure}
\centering
\includegraphics[width=7.5cm]{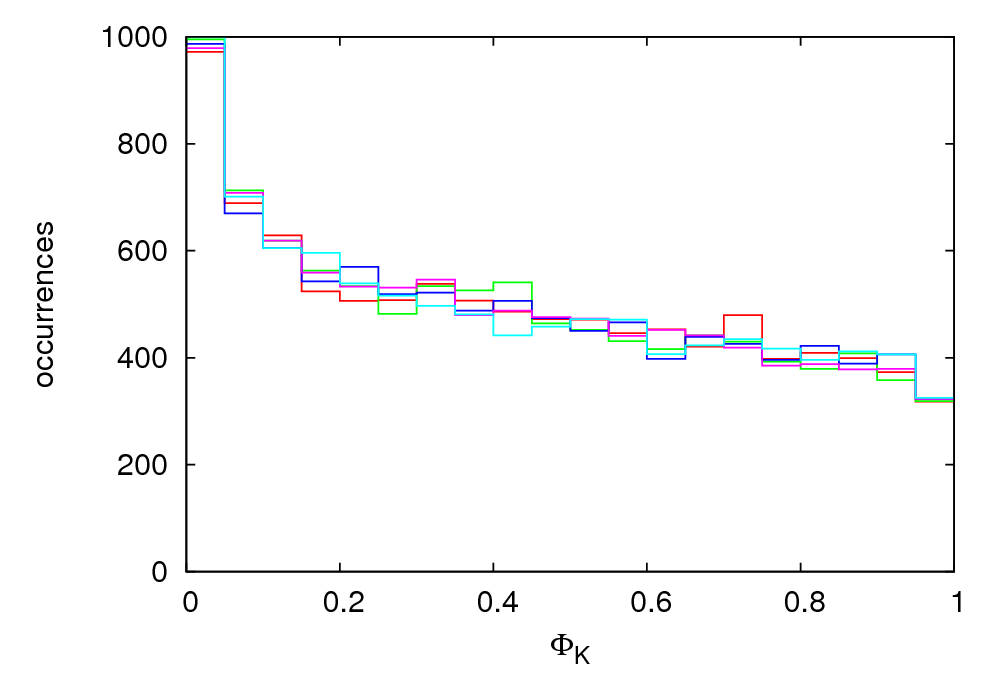}
\includegraphics[width=7.5cm]{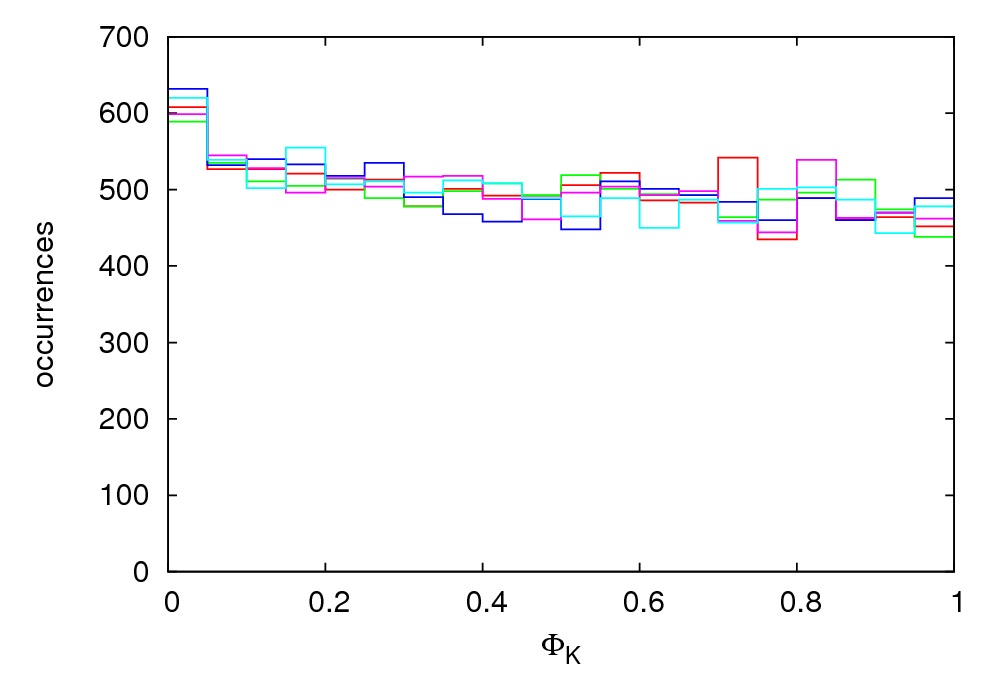}
\includegraphics[width=7.5cm]{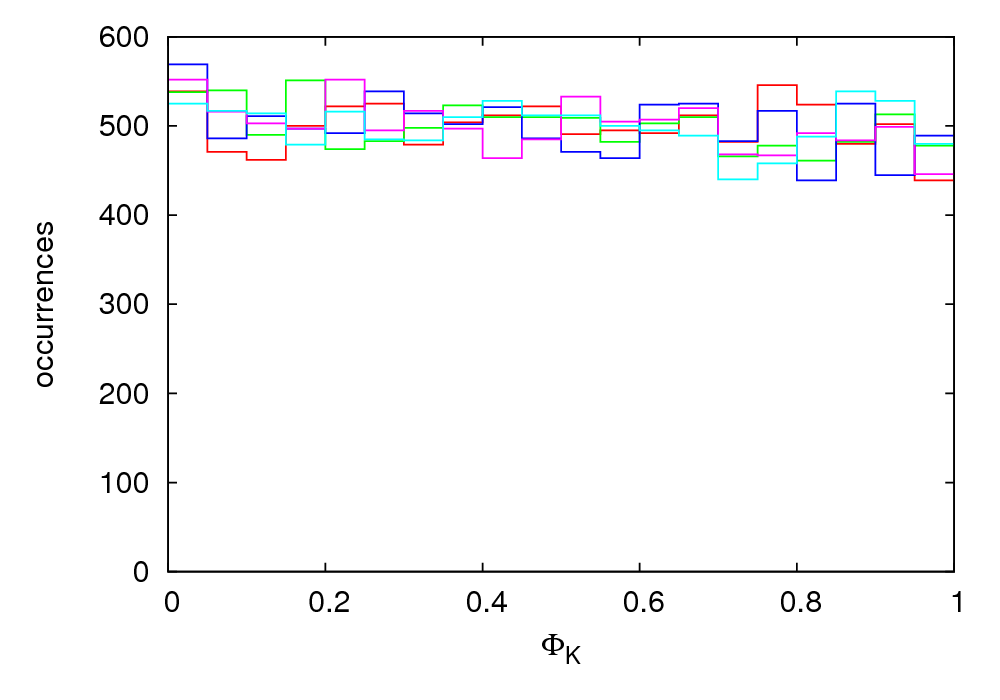}
\caption{Effects from small $n$ on the histogram of $\Phi_K$. \emph{Upper left:} $n=10$, \emph{upper right:} $n=143$, \emph{bottom:} $n=2000$. For all histograms, we used $m=10000$, and we show 5 realizations. For $n \rightarrow \infty$, we expect the histogram to become flat.}
\label{f:hist_bias}
\end{figure}

\begin{figure}
\centering
\includegraphics[width=7.5cm]{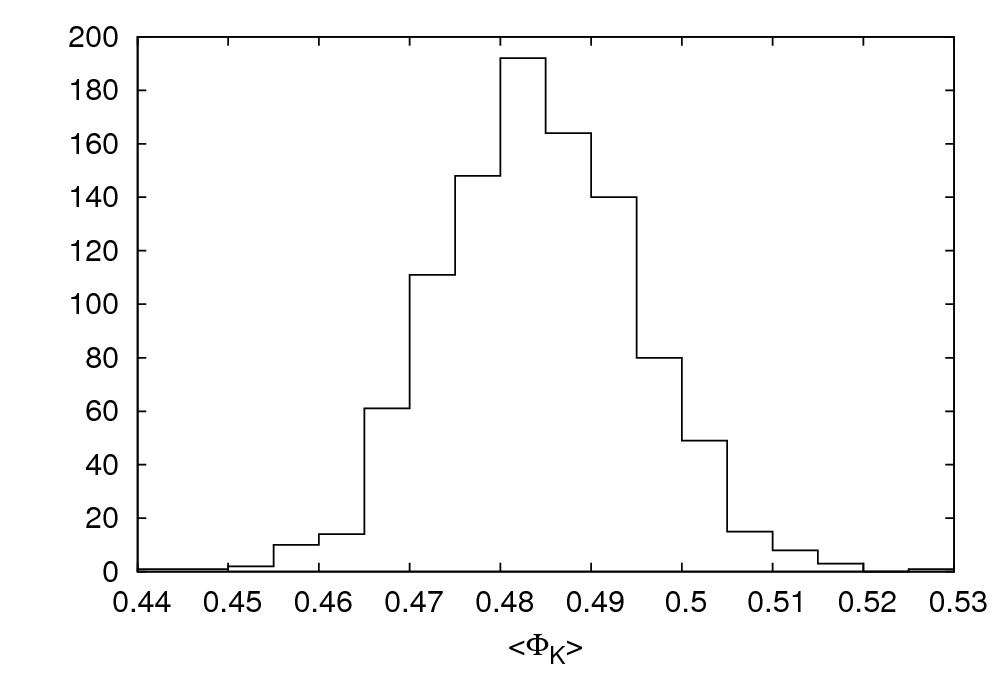}
\includegraphics[width=7.5cm]{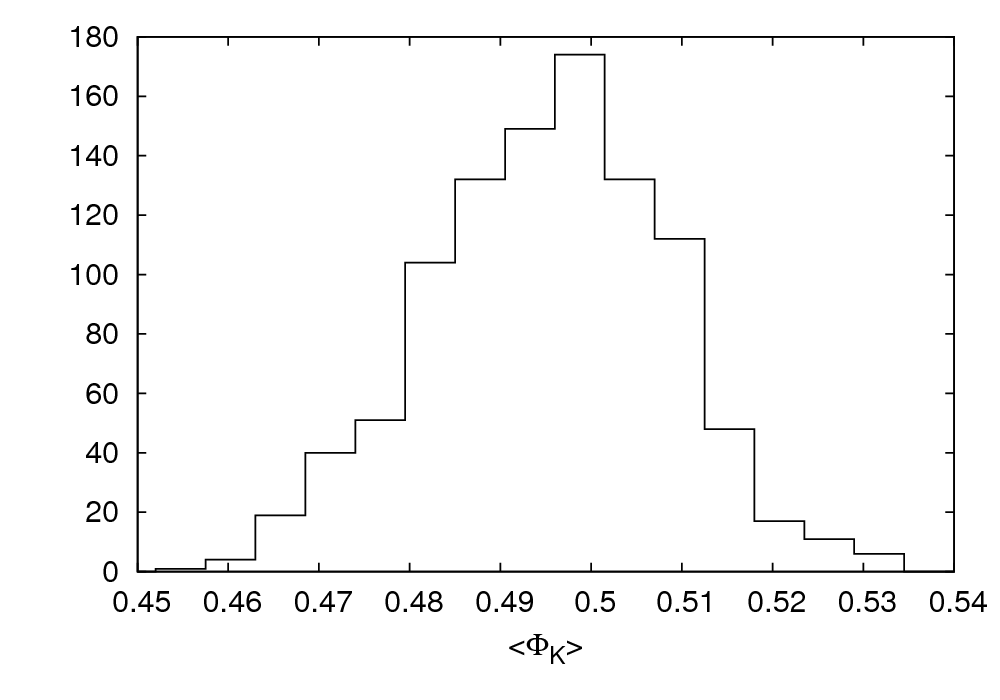}
\caption{Histogram of $\overline{\Phi_K}$ computed from 1000 realizations. \emph{Left panel:} $n=143$, $m=712$. The histogram is centered around $0.484$, which means that $\overline{\Phi_K}$ is biased low by about $1.5\,\sigma$. \emph{Right panel:} $n=2000$, $m=500$. The histogram is centered around $0.495$, the bias is less than $0.5 \, \sigma$.}
\label{f:bias_hist}
\end{figure}

In Section \ref{sec:effects}, we have looked at the effects from small $n$ on the partition function of $\la_n$, and we found that the the partition function for $n \approx 100$ is already very close to the Kolmogorov-function. 
In this Appendix, we study the effect from small $n$ in more detail. 

In Fig.~\ref{f:bias}, we compare simulations of the empirical partition function of $\la_n$ for $n=143$ with the one for $n=2000$. In both plots, we used $m=712$, which is what we used in the local analysis of WMAP data using ring segments, and we show 100 realizations of the partition function. For $n=143$, most of the empirical partition functions lie to the left of the Kolmogorov-curve. 

The same effect can be seen in Fig.~\ref{f:hist_bias}, where we show a histogram of all $\Phi_K$ for $n=10$ (upper left), $n=143$ (upper right), and $n=2000$ (bottom panel). Here, we have used $m=10000$ in order to minimize the intrinsic variation of the histogram, and we show 5 realizations. Without finite-size effects, these histograms should be flat, as we argued in Section \ref{s:decorr}. The finite-size effect causes a bias of $\Phi_K$ towards smaller values.

In order to quantify the resulting bias of $\overline{\Phi_K}$ for our local analysis in Section \ref{s:decorr} $(n=143)$, we compute $\overline{\Phi_K}$ for $n=143$ and $m=712$ for 1000 samples, and plot the resulting histogram of $\overline{\Phi_K}$ in the left panel of Fig.~\ref{f:bias_hist}. We do the same thing for the global analysis in Section \ref{s:global} for $n=2000$ and $m=500$, and plot the resulting histogram of $\overline{\Phi_K}$ in the right panel of Fig.~\ref{f:bias_hist}.

Without the finite-size effects, these should be Gaussian distributed around $\langle \Phi_K \rangle = 0.5$ with a standard deviation of $\sigma = 1/\sqrt{12m} \approx 0.011$ $(0.013)$ for $m=712$ ($m=500$), as we have explained in Section \ref{s:decorr}. The histogram for $n=143$ is centered around $0.484$ rather than around $0.5$, i.e.\ $\overline{\Phi_K}$ is biased low by about $1.5 \,\sigma$. The histogram for $n=2000$  is centered around $0.495$, i.e.\ the bias is less than $0.5\,\sigma$.

An analogous analysis yields that for $n=256$, $\overline{\Phi_K}$ is distributed around $0.489$. We used this in Section \ref{s:decorr} when comparing the results of the de-correlation method with theoretical expectations.

\section{Residual correlations between ring segments and global samples}\label{s:res_corr}

\begin{figure}
\centering
\includegraphics[width=7.5cm]{figs/scatter_of_X_around_Phi_K_n_143_m_712.png}
\includegraphics[width=7.5cm]{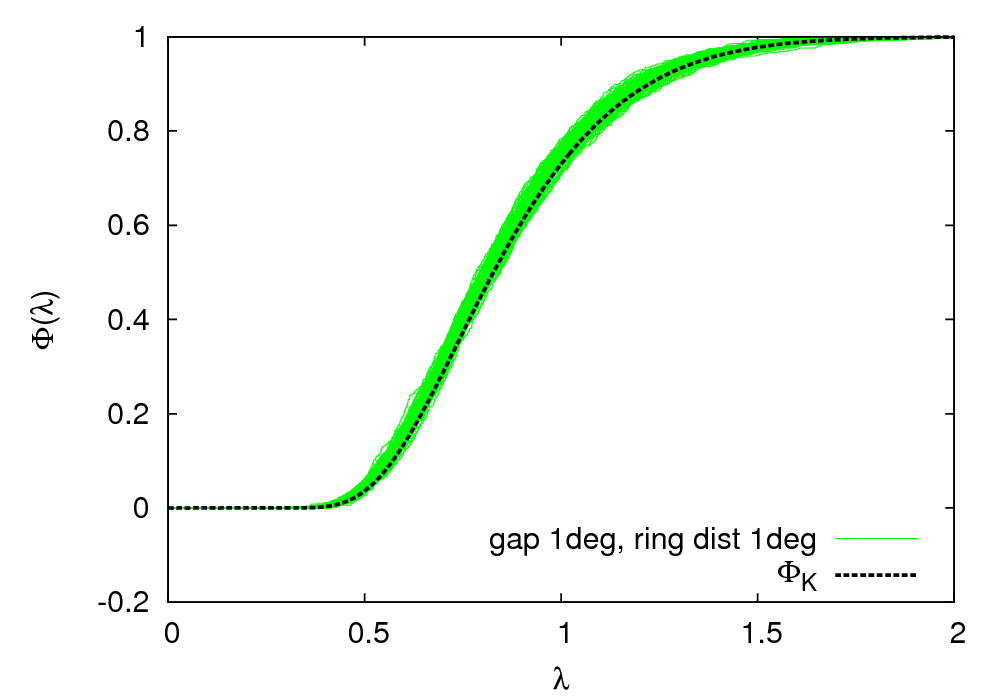}
\includegraphics[width=7.5cm]{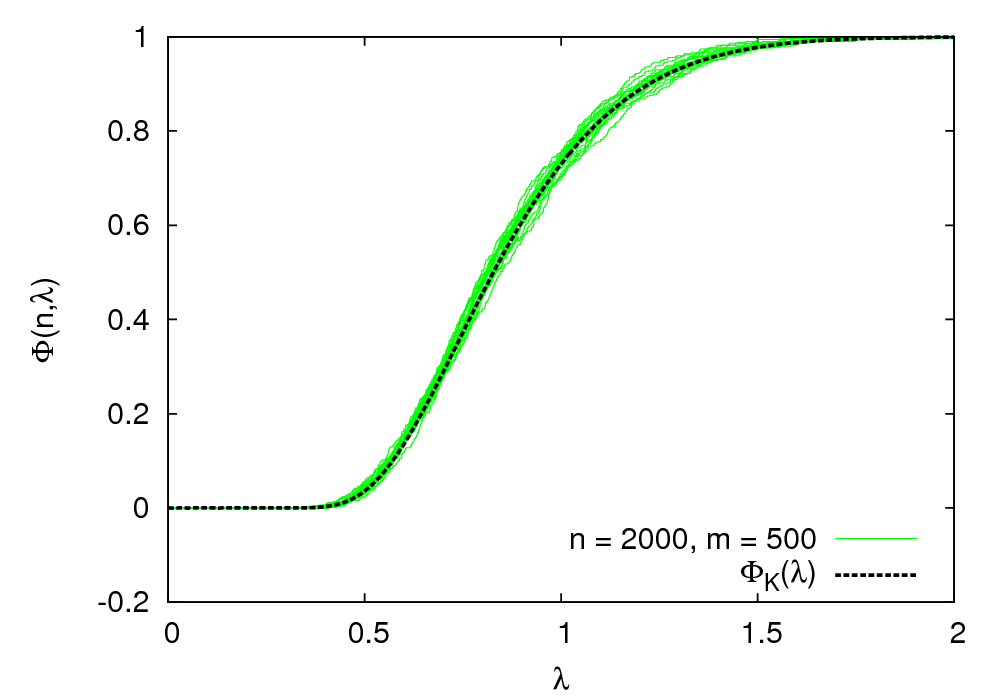}
\includegraphics[width=7.5cm]{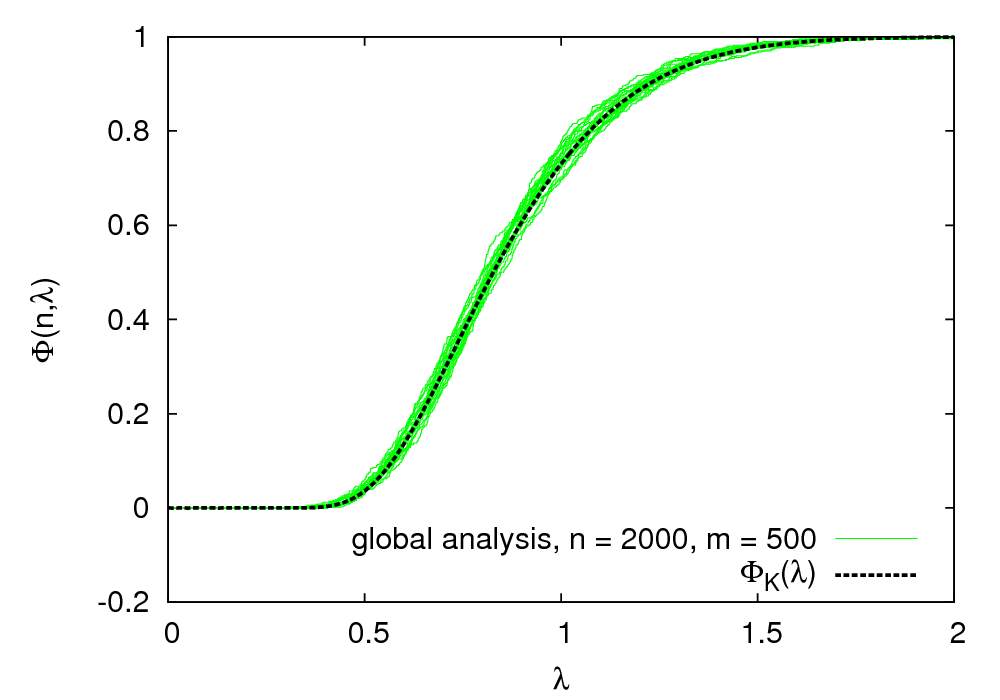}
\caption{Investigation of effects from correlations between ring segments (\emph{top}) and between samples of the global analysis (\emph{bottom}). In the top (bottom) panels, we show 100 (20) realizations of the simulated empirical partition function of $\la_n$ obtained from the following samples: \newline
\emph{Top left:} $m=712$ independent samples of $n=143$ pixels each. \newline
\emph{Top right:} $m=712$ ring segments of an opening angle of $20^\circ$ ($n \approx 143$), and a gap between rings and ring segments of $1^\circ$, extracted from simulated CMB maps. \newline
\emph{Bottom left:} $m=500$ independent samples of $n=2000$ pixels each. \newline
\emph{Bottom right:} $m=500$ global samples of $n=2000$ pixels each, randomly drawn from simulated CMB maps.}
\label{f:res_corr}
\end{figure}

In Fig.~\ref{f:res_corr}, we study the effects of residual correlations between the ring segments and between the global samples. One would expect that these correlations cause the partition functions of $\la_n$ to be too steep as we have argued above. However, in Fig.~\ref{f:res_corr}, we verify that for our analyses we do not yet see any effect of this.

\end{appendix}

\end{document}